\documentclass[prd,preprint,superscriptaddress,tightenlines,nofootinbib,
eqsecnum,showpacs]{revtex4-1}

\usepackage{amsmath}
\usepackage{amsfonts}
\usepackage{amssymb}
\usepackage{bm}
\usepackage{hyperref}
\usepackage{mathrsfs}
\usepackage{graphicx}
\usepackage{color}

\allowdisplaybreaks
\DeclareSymbolFontAlphabet{\mathrsfs}{rsfs}
\DeclareMathAlphabet{\mathcal}{OMS}{cmsy}{m}{n}

\newcommand{\ud}{\mathrm{d}}
\newcommand{\ui}{\mathrm{i}}
\newcommand{\beq}{\begin{equation}}
\newcommand{\eeq}{\end{equation}}

\begin{document}

\title{The third and a half post-Newtonian gravitational wave quadrupole mode
  for quasi-circular inspiralling compact binaries}

\author{Guillaume Faye}\email{faye@iap.fr}
\affiliation{$\mathcal{G}\mathbb{R}\varepsilon{\mathbb{C}}\mathcal{O}$,
  Institut d'Astrophysique de Paris --- UMR 7095 du CNRS, \\ Universit\'e
  Pierre \& Marie Curie, 98\textsuperscript{bis} boulevard Arago, 75014 Paris,
  France}

\author{Sylvain Marsat}\email{marsat@iap.fr}
\affiliation{$\mathcal{G}\mathbb{R}\varepsilon{\mathbb{C}}\mathcal{O}$,
  Institut d'Astrophysique de Paris --- UMR 7095 du CNRS, \\ Universit\'e
  Pierre \& Marie Curie, 98\textsuperscript{bis} boulevard Arago, 75014 Paris,
  France}

\author{Luc Blanchet}\email{blanchet@iap.fr}
\affiliation{$\mathcal{G}\mathbb{R}\varepsilon{\mathbb{C}}\mathcal{O}$,
  Institut d'Astrophysique de Paris --- UMR 7095 du CNRS, \\ Universit\'e
  Pierre \& Marie Curie, 98\textsuperscript{bis} boulevard Arago, 75014 Paris,
  France}

\author{Bala R. Iyer} \email{bri@rri.res.in} 
\affiliation{Raman Research Institute,
Bangalore 560 080, India} 

\date{\today}

\begin{abstract}
  We compute the quadrupole mode of the gravitational waveform of inspiralling
  compact binaries at the third and half post-Newtonian (3.5PN) approximation
  of general relativity. The computation is performed using the multipolar
  post-Newtonian formalism, and restricted to binaries without spins moving on
  quasi-circular orbits. The new inputs mainly include the 3.5PN terms in the
  mass quadrupole moment of the source, and the control of required
  subdominant corrections to the contributions of hereditary integrals (tails
  and non-linear memory effect). The result is given in the form of the
  quadrupolar mode $(2,2)$ in a spin-weighted spherical harmonic decomposition
  of the waveform, and may be used for comparison with the counterpart
  quantity computed in numerical relativity. It is a step towards the
  computation of the full 3.5PN waveform, whose knowledge is expected to
  reduce the errors on the location parameters of the source.
\end{abstract}


\maketitle

\section{Introduction}\label{sec:intro}

The first ever direct detection of gravitational waves might occur around
2016, when the current network of ground-based laser-interferometric detectors
(such as VIRGO and LIGO) will have been upgraded to a higher sensitivity. 
Inspiralling and merging binary systems composed of black holes and/or neutron
stars are among the most promising sources for those detectors. Though the
gravitational waves are extremely weak, the large number of predictable cycles
in the detector's bandwidth will enable the on-line detection and later the
parameter estimation of the signal. The detector output will be
cross-correlated with a number of copies of the theoretically predicted signal
or template, corresponding to different signal parameters. The need of a
faithful template bank has driven the development over the last twenty years
of both accurate approximation methods and powerful numerical schemes in
general relativity. For a review of gravitational wave detectors and future
gravitational wave astronomy, see~\cite{sathyaschutz09}.

The templates for the inspiral and merger of two compact objects (say black
holes) are computed by combining a high order post-Newtonian approximation for
the early inspiral phase~\cite{Bliving}, with a full-fledged numerical
integration of the field equations for the late inspiral and ringdown
phases~\cite{Pretorius,Baker,Campanelli,BCPZ,Hannam09}. The post-Newtonian and
numerically-generated results are then matched together with high precision,
yielding the full gravitational waveform, including all amplitude and phase
modulations, described either analytically and/or
numerically~\cite{BCP07,Berti,Jena,Boyle,Ajith08,Buo09,Pan09,PanBFRT11}.

Previous post-Newtonian works~\cite{BIWW96,ABIQ04,KBI07,K08,BFIS08} have
provided the waveform including all its harmonics (\textit{i.e.} beyond the
dominant harmonic at twice the orbital frequency) up to the 3PN
order.\footnote{As usual we refer to $n$PN as the post-Newtonian terms with
  formal order $\mathcal{O}(c^{-2n})$ relative to the Newtonian acceleration
  in the equations of motion, or to the quadrupole-moment formalism for the
  radiation field.} In applications to data analysis the full waveform should
be used for detection and parameter estimation up to the maximum available
post-Newtonian order. In particular, it should further improve the angular
resolution and the distance measurement of the system for massive enough
binaries \cite{AISSV,TS08}.
Now, the quadrupole mode at twice the orbital
frequency, having $(\ell,m)=(2,2)$ in a spin-weighted spherical harmonic
decomposition, is the dominant one in the sense that it gives the only
time-varying contribution to the waveform at the dominant order.\footnote{The
  $(\ell,m)=(2,0)$ mode contributes at the dominant Newtonian order, but only
  in the form of a non-oscillating (DC) term.} It is also the one which is
computed with the best precision in most numerical simulations.
In the present paper we extend the previous
works~\cite{BIWW96,ABIQ04,KBI07,K08,BFIS08} by computing the dominant $(2,2)$
mode at the next 3.5PN order. Our result agrees with the one derived within
the black-hole perturbation theory in the limit where the binary mass
ratio goes to zero \cite{TSasa94,FI10}. The completion of the other modes at
the 3.5PN order does not appear to be straightforward, notably regarding the
$(2,1)$ mode, and will be left for future investigation.

The computational basis is the multipolar post-Newtonian wave-generation
formalism, which has two different aspects. First, it constitutes a general
method applicable to extended sources with compact support, which combines a
mixed post-Minkowskian and multipolar expansion for the field outside the
source~\cite{BD86,BD92,B98quad,B98tail}, with a post-Newtonian expansion for
the field inside the source~\cite{B98mult}. Second, it addresses the problem of
applying this method to point-particles modelling compact
objects~\cite{BIJ02,BI04mult}, in which case it crucially requires a self-field
regularization~\cite{BFreg}. In the present article
we shall mainly focus on the results and refer to previous papers for full
details on this formalism (see also~\cite{BFIS08} for a summary).\footnote{In
  the following we shall refer to Ref.~\cite{BFIS08} as
  Paper I.}
 
Our plan will be as follows. In Sect.~\ref{sec:eom} we recall the equations of
motion of non-spinning compact binaries on quasi-circular orbits up to 3.5PN
order. In Sect.~\ref{sec:gwf} we remind some basic definitions for computing
the gravitational waveform and associated polarization modes for planar
compact binaries. In Sect.~\ref{sec:quadmode} we express the dominant
radiative quadrupole moment in terms of the source multipole moments of a
general isolated source up to 3.5PN order. In Sect.~\ref{sec:sourcemoments} we
give the expressions of those source moments fully reduced in the case of
circular compact binaries at the same accuracy level. Finally the result for
the $(2,2)$ mode at 3.5PN order is presented in Sect.~\ref{sec:h22}. The
source quadrupole moment at 3.5PN order for non-circular binaries in a
center-of-mass frame is relegated to Appendix~\ref{sec:appA}.

\section{Quasi-circular binary at 3.5PN order}\label{sec:eom}

An inspiralling compact binary of non-spinning compact bodies is modelled as a
system of two particles solely described by their masses $m_1$ and $m_2$. The
orbital plane is spanned by the relative position of the particles $\bm{x} =
\bm{y}_1 - \bm{y}_2$ and the relative ordinary velocity $\bm{v} =
\ud\bm{x}/\ud t$. The unit vector normal to this plane is given by
$\bm{\ell}=\bm{x}\times\bm{v}/\vert\bm{x}\times\bm{v}\vert$ (we assume a
non-radial orbit) and is constant in the absence of spin effects. Introducing
the unit separation direction $\bm{n}=\bm{x}/r$, where $r=\vert\bm{x}\vert$ is
the separation distance, and posing $\bm{\lambda} = \bm{\ell}\times\bm{n}$ we
have
\begin{subequations}\label{eq:va}\begin{align}
\bm{x} &= r\bm{n}\,,\label{eq:x}\\
\bm{v} &= \dot r \,\bm{n}
+ r \,\omega \,\bm{\lambda}\,,\label{eq:v}\\ 
\frac{\ud \bm{v}}{\ud t} &= \left(\ddot{r} - r
\,\omega^2\right) \,\bm{n} + \left(r \,\dot{\omega} + 2 \dot{r} \,\omega\right)
\,\bm{\lambda}\,,\label{eq:a}
\end{align}\end{subequations}
which defines the orbital frequency $\omega$ related in the usual way to the
orbital phase $\varphi$ by $\omega = \dot \varphi$. The motion of the binary
follows a quasi-circular orbit decaying by the effect of radiation
reaction starting at 2.5PN order. Using the facts that
$\dot{r}=\mathcal{O}(c^{-5})$ and $\dot{\omega}=\mathcal{O}(c^{-5})$, and
noticing that $\ddot{r}=\mathcal{O}(c^{-10})$ is of the order of the square of
radiation reaction effects, we obtain the quasi-circular acceleration as
\begin{equation}\label{eq:acirc}
\frac{\ud \bm{v}}{\ud t} = - \omega^2 \,\bm{x} +
\Bigl(\frac{\dot{\omega}}{\omega} +
2\frac{\dot{r}}{r}\Bigr)\,\bm{v}\,+\mathcal{O}\left(\frac{1}{c^{10}}\right)\,.
\end{equation}

The conservative part of the dynamics is given for circular orbits by the
expression of the orbital frequency $\omega$ in terms of the binary's
separation $r$ up to 3PN order. This result has been obtained in harmonic
coordinates~\cite{BF00,BFeom,BDE04,IFA01,itoh1,itoh2,FS11} and in
Arnowitt-Deser-Misner (ADM) coordinates~\cite{JaraS98,JaraS99,DJSdim}. For the
present work $r$ is the orbital separation in harmonic coordinates, and from
it we define the post-Newtonian parameter
\begin{equation}\label{eq:PNgam}
\gamma=\frac{G m}{r c^2}\,.
\end{equation}
Our mass parameters will be the total mass $m=m_1+m_2$ and the symmetric mass
ratio $\nu=m_1m_2/m^2$. The orbital frequency is then given by the 3PN
``Kepler's law''
\begin{align}\label{eq:omega3PN}
\omega^2 &= {G m\over r^3}\biggl\{ 1+\gamma\Bigl(-3+\nu\Bigr) + \gamma^2
\left(6+\frac{41}{4}\nu +\nu^2\right) \nonumber\\ &~~\qquad+\gamma^3
\left(-10+\left[-\frac{75707}{840}+\frac{41}{64}\pi^2
+22\ln\left(\frac{r}{r'_0}\right) \right]\nu +\frac{19}{2}\nu^2+\nu^3\right)
\biggr\}\,.
\end{align}
We neglect the 4PN and higher terms $\mathcal{O}(\gamma^4)$. The logarithm at
3PN order comes from a Hadamard self-field regularization scheme
\cite{BF00,BFeom} and involves a regularization constant $r'_0$ specific to
harmonic coordinates. This constant $r'_0$ will disappear from our physical
results in the end. By inverting \eqref{eq:omega3PN} we obtain $\gamma$ in
terms of the alternative parameter
\begin{equation}\label{eq:PNx}
x = \left({G\,m\,\omega \over c^3}\right)^{2/3}\,,
\end{equation}
which is an invariant in a large class of coordinate systems\footnote{Those
  are the coordinates for which the metric is asymptotically Minkowskian far
  from the source.} including for instance the harmonic and ADM coordinates.
At 3PN order we have
\begin{align}\label{eq:gammax}
\gamma &= x \biggl\{ 1+x \left(1-\frac{\nu}{3}\right) + x^2
 \left(1-\frac{65}{12}\nu\right) \nonumber\\ &~~\qquad+ x^3
 \left(1+\left[-\frac{2203}{2520}-\frac{41}{192}\pi^2
 -\frac{22}{3}\ln\left(\frac{r}{r'_0}\right) \right]\nu
 +\frac{229}{36}\nu^2+\frac{\nu^3}{81}\right) \biggr\}\,.
\end{align}

The dissipative radiation reaction part of the equations of motion can be
computed by balancing the change in the orbital energy $E$ with the total
energy flux $\mathcal{F}$ radiated by the gravitational waves, $\ud E/\ud
t=-\mathcal{F}$. Up to 3.5PN order, which means using the orbital energy and
energy flux at 1PN relative order, this gives
\begin{equation}\label{eq:rdot}
\dot{r} = - \frac{64}{5} \sqrt{\frac{G m}{r}}~\nu\,\gamma^{5/2}\left[1+ \gamma
  \left(-\frac{1751}{336} - \frac{7}{4}\nu\right)\right]\,.
\end{equation}
Using the post-Newtonian law \eqref{eq:omega3PN} truncated at 1PN order we
further deduce
\begin{equation}\label{eq:omdot}
\dot{\omega} =
\frac{96}{5} \,\frac{G m}{r^3}\,\nu\,\gamma^{5/2}\left[1+ \gamma
  \left(-\frac{2591}{336} - \frac{11}{12}\nu\right)\right]\,.
\end{equation}
These expressions can be obtained alternatively using the center-of-mass 3.5PN
equations of motion for general orbits \cite{PW02,NB05}. By substituting them
into Eq.~\eqref{eq:acirc} we obtain the complete acceleration for
quasi-circular orbits at 3.5PN order as
\begin{equation}\label{eq:acircfinal}
\frac{\ud \bm{v}}{\ud t} = - \omega^2 \,\bm{x} - \frac{32}{5}\,\sqrt{\frac{G
m}{r^3}}\,\,\nu\,\gamma^{5/2}\left[1+ \gamma \left(-\frac{743}{336} -
\frac{11}{4}\nu\right)\right]\bm{v} \,,
\end{equation}
where the radiation reaction force up to 3.5PN order is explicitly exhibited.
As for the velocity it is given with the same accuracy by
\begin{equation}\label{eq:vcircfinal}
\bm{v} = r \,\omega \,\bm{\lambda} - \frac{64}{5}\,\sqrt{\frac{G
m}{r}}\,\,\nu\,\gamma^{5/2}\left[1+ \gamma \left(-\frac{1751}{336} -
\frac{7}{4}\nu\right)\right]\bm{n} \,.
\end{equation}

\section{Gravitational waveform of non-spinning binaries}\label{sec:gwf}

\subsection{General definitions}

The gravitational waveform propagating in the asymptotic regions of an
isolated source is defined in a radiative (Bondi-type) coordinate system
$X^\mu=(c T,\bm{X})$ by \cite{Th80}
\begin{align}\label{eq:hij}
h_{ij}^\text{TT} &= \frac{4G}{c^2R} \,\mathcal{P}_{ijkl}
(\bm{N}) \sum^{+\infty}_{\ell=2}\frac{1}{c^\ell \ell !} \biggl\{ N_{L-2}
\,U_{klL-2}(T_R) - \frac{2\ell}{c(\ell+1)} \,N_{aL-2} \,\varepsilon_{ab(k}
\,V_{l)bL-2}(T_R)\biggr\} \,,
\end{align}
where terms of order $\mathcal{O}(1/R^2)$ are neglected. Here
$R=\vert\bm{X}\vert$ and $\bm{N}=\bm{X}/R$ are the distance and the direction
of the source. The Minkowskian retarded time is denoted $T_R=T-R/c$. The
transverse-tracefree (TT) projection operator reads $\mathcal{P}_{ijkl} =
\mathcal{P}_{i(k}\mathcal{P}_{l)j}-\frac{1}{2}\mathcal{P}_{ij}\mathcal{P}_{kl}$
where $\mathcal{P}_{ij}=\delta_{ij}-N_iN_j$ is the projector orthogonal to the
unit direction $\bm{N}$. The waveform \eqref{eq:hij} is parametrized by two
sets of radiative symmetric-trace-free (STF) multipole moments, of mass-type,
$U_L$, and of current-type, $V_L$, which are functions of the retarded time
$T_R$. The notation for STF tensors and multi-indices such as $L=i_1\cdots
i_\ell$ (where $\ell$ denotes the multipole order) is the same as in
Ref.~\cite{BFIS08} (Paper I) where it is fully detailed.

\begin{figure}
	\includegraphics[width=7cm]{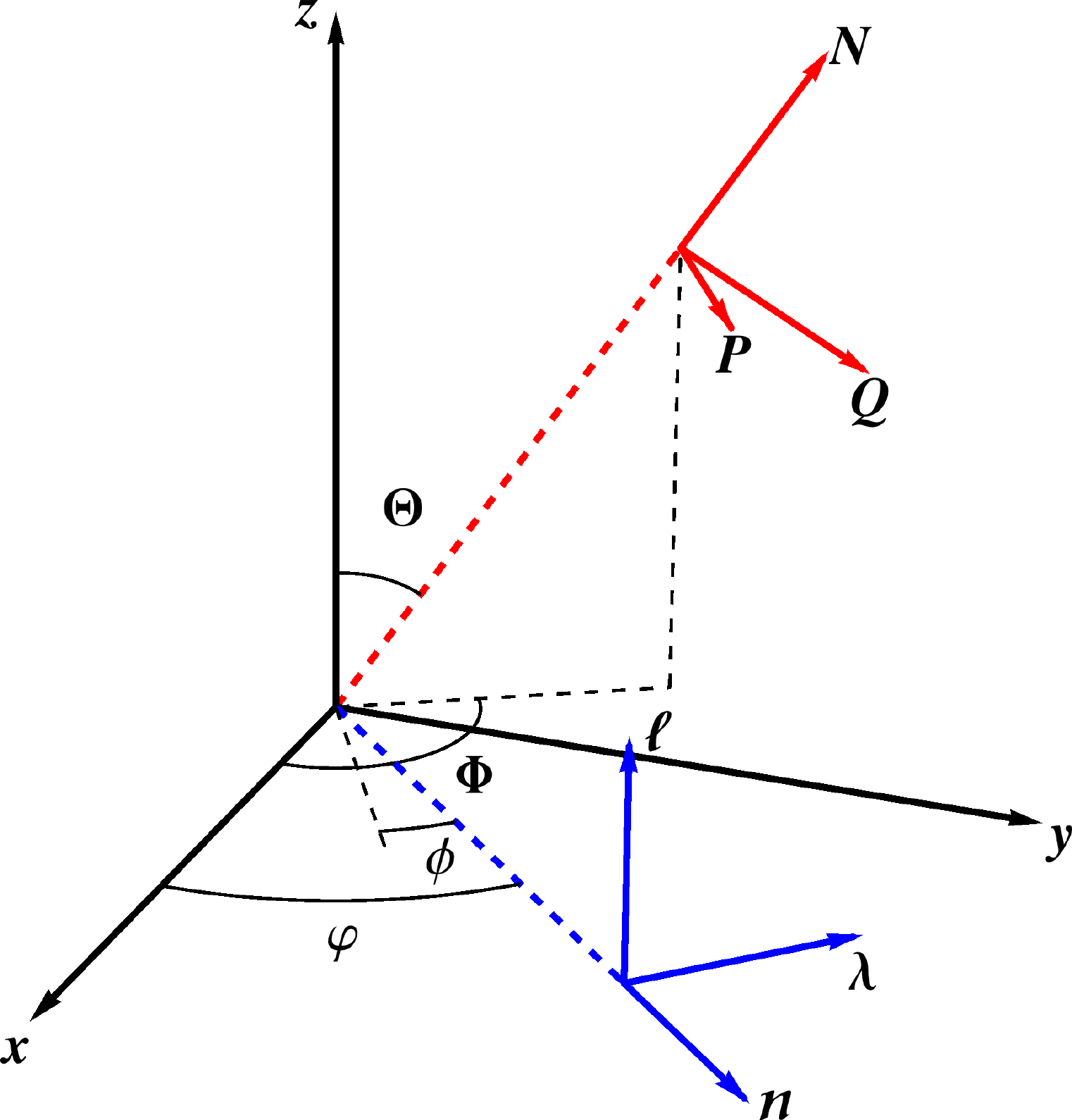}
	\caption{Definition of angles and polarization vectors. $\bm{N}$ points
      towards the observer, whose spherical coordinates are $(\Theta,\Phi)$,
      with $\Theta=\pi/2$ corresponding to the orbital plane of the binary;
      $\Theta$ is also the inclination of the orbit on the sky from the
      observer's viewpoint. We define then $\bm{P}=-\bm{e}_\Phi$ and
      $\bm{Q}=\bm{e}_\Theta$. The ($\bm{n}$, $\bm{\lambda}$,
      $\bm{\ell}$) triad is displayed, $\varphi$ being the orbital phase, and
      $\phi \equiv 
      \varphi-\Phi+\pi/2$ being the orbital phase relative to the
      direction of the ascending node of the orbit, which is also the
      direction of $\bm{P}$.}\label{fig:frame}
\end{figure}

We now introduce two unit polarisation vectors $\bm{P}$ and $\bm{Q}$ forming
with $\bm{N}$ an orthonormal triad, as indicated on Fig.~\ref{fig:frame}.
Representing the observers's direction in spherical coordinates
$(\Theta,\Phi)$, with $\Theta=\pi/2$ corresponding to the orbital plane of the
binary, our choice for the polarization vectors (which is not
unique\footnote{Beware that other authors use different conventions for the
  polarization triad. In Ref.~\cite{K08}, this results in an overall sign
  difference with us for both $h_+$ and $h_\times$.}) is $\bm{P}=-\bm{e}_\Phi$
and $\bm{Q}=\bm{e}_\Theta$, where $\bm{e}_\Theta = \partial
\bm{N}/\partial \Theta$ and $\bm{e}_\Phi=\partial \bm{N}/(\sin
\Theta \partial \Phi)$ are the standard transverse basis vectors of spherical
coordinates. With respect to the polarization vectors $\bm{P}$ and $\bm{Q}$,
the two polarization states are

\begin{subequations}\label{eq:hpc}\begin{align}
h_+ &= \frac{1}{2}\left(P_iP_j-Q_iQ_j\right) h_{ij}^\text{TT}\,,\\
h_\times &= \frac{1}{2}\left(P_iQ_j+P_jQ_i\right) h_{ij}^\text{TT}\,.
\end{align}\end{subequations}
We shall henceforth pose $h \equiv h_+ -\ui h_\times$. For quasi-circular
orbits the complex wave-amplitude $h$ will be a function of the orbital phase
$\varphi=\int\omega\,\ud t$ and of the spherical angles $(\Theta,\Phi)$
associated with $\bm{N}$. Next, we apply a decomposition in terms of
spin-weighted spherical harmonics $Y^{\ell m}_{-2}(\Theta,\Phi)$.\footnote{See
  Ref.~\cite{K08}; our convention for the spin-weighted spherical harmonics is
  the same as in Paper I.} This yields the definition of the spherical modes
$h^{\ell m}$ of the waveform as
\begin{equation}\label{eq:hdecomp}
h = \sum^{+\infty}_{\ell=2}\sum^{\ell}_{m=-\ell} h^{\ell
m} \,Y^{\ell m}_{-2}(\Theta,\Phi)\,.
\end{equation}
From the orthogonality properties of the $Y^{\ell m}_{-2}$'s,
the separate modes $h^{\ell m}$ are extracted from $h$ by the surface integral
(with solid angle element $\ud^2 \Omega=\sin\Theta\ud\Theta\ud\Phi$)
\begin{equation}\label{eq:hlm}
h^{\ell m} = \int \ud^2\Omega \,h \,\overline{Y}^{\,\ell
m}_{-2} (\Theta,\Phi)\,,
\end{equation}
where the overbar means the complex conjugation. We can thus insert the
waveform decomposition \eqref{eq:hij} in terms of STF radiative moments and
perform the angular integration, to obtain
\begin{equation}\label{eq:inv}
h^{\ell m} = -\frac{G}{\sqrt{2}\,R\,c^{\ell+2}}\left[U^{\ell
m}-\frac{\ui}{c}V^{\ell m}\right]\,,
\end{equation}
where $U^{\ell m}$ and $V^{\ell m}$ are the radiative mass and current moments
in standard (non-STF) form \cite{Th80,K08}, related to the STF radiative
moments by
\begin{subequations}\label{eq:UV}\begin{align}
U^{\ell m} &= \frac{4}{\ell!}\,\sqrt{\frac{(\ell+1)(\ell+2)}{2\ell(\ell-1)}}
\,\alpha_L^{\ell m}\,U_L\,,\\ V^{\ell m} &=
-\frac{8}{\ell!}\,\sqrt{\frac{\ell(\ell+2)}{2(\ell+1)(\ell-1)}}
\,\alpha_L^{\ell m}\,V_L\,.
\end{align}\end{subequations}
Here we have introduced the tensor $\alpha_L^{\ell m}$ defined by the
decomposition of the STF product of $\ell$ unit vectors $\bm{N}$,
\textit{i.e}. $\hat{N}_L$ in our notation, into ordinary spherical harmonics,
namely
\begin{equation}\label{eq:alphaL}
\hat{N}_L = \sum_{m=-\ell}^{\ell}
\alpha_L^{\ell m}\,Y^{\ell m}(\Theta,\Phi)\,.
\end{equation}

\subsection{Mode separation for planar binaries}

Let us now show that for planar binaries, the mode $h^{\ell m}$ is in fact
entirely given by the ``mass'' contribution [first term in \eqref{eq:inv}]
when $\ell+m$ is even, and by the ``current'' contribution [second term in
\eqref{eq:inv}] when $\ell+m$ is odd. Notice that this is valid in general for
non-spinning binaries, regardless of the orbit being quasi-circular or
elliptical, or for spinning binaries with spins aligned or anti-aligned with
the orbital angular momentum. The important point is only that the black holes
must be non-precessing so that the motion is planar. This mode separation has
been already used in Ref.~\cite{K08}, but we explain here how to deduce it
from an argument of parity invariance. If an observer with space-coordinates
$\bm{X}$ in a spatial grid $\{X^i\}$ looks at the gravitational-wave signal at
a given time $T$, he finds that the source particles lie at some positions
$\bm{y}_A$ and possess certain velocities $\bm{v}_A$, for $A=1,2$. Now, a
second observer located at $\bm{X}'= - \bm{X}$ and looking at a
parity-reversed system with positions $\bm{y}'_A=-\bm{y}_A$ and velocities
$\bm{v}'_A=-\bm{v}_A$ must observe the same waveform $h_{ij}^\text{TT}$, by
parity invariance of the Einstein field equations. This translates
into:\footnote{In the non-precessing case, the aligned or anti-aligned spins are
only characterized by their magnitude, which is a constant parameter that does
not play a role here.}
\begin{equation} \label{eq:parity_hTT}
h^\text{TT}_{ij}(-\bm{X}, T; -\bm{y}_A, -\bm{v}_A) =
h^{\text{TT}}_{ij}(\bm{X}, T; \bm{y}_A, \bm{v}_A)\, .
\end{equation} 
Because the multipolar decomposition \eqref{eq:hij} is unique, this result
allows us to recover the basic built-in parity properties of the mass and
current moments $U_L$ and $V_L$, regarded as functionals of the matter
variables $\bm{y}_A$, $\bm{v}_A$:
\begin{subequations}
\begin{align}
& U_L(T;-\bm{y}_A,-\bm{v}_A) = (-1)^\ell  \,U_L(T;\bm{y}_A,\bm{v}_A) \, , \\
& V_L(T;-\bm{y}_A,-\bm{v}_A) = (-1)^{\ell+1} \,V_L(T;\bm{y}_A,\bm{v}_A) \, .
\end{align}
\end{subequations}
To see how Eq.~\eqref{eq:parity_hTT} reflects on the two polarizations, we
remark that the polarization vector
$\bm{P}=\bm{N}\times\bm{\ell}/\vert\bm{N}\times\bm{\ell}\vert$ transforms as
$\bm{P}'=-\bm{P}$, while $\bm{Q}=\bm{N}\times\bm{P}$ is clearly invariant, due
to the fact that $\bm{\ell}$ is itself invariant and $\bm{N}'=-\bm{N}$.
Those considerations applied to the complex polarization $h = h_+ - \ui
h_\times$, with the definition~\eqref{eq:hpc} for $h_+$ and $h_\times$, show
that, in the center-of-mass frame
\begin{equation} \label{eq:parity_h}
h(-\bm{X}, T; -\bm{x}, -\bm{v}) = \overline{h}(\bm{X}, T; \bm{x}, \bm{v})\, ,
\end{equation} 
where $\bm{x} = \bm{y}_1 - \bm{y}_2$ and $\bm{v} = \ud\bm{x}/\ud t$.

For planar motions, it is particularly convenient to rewrite the previous
relations in the spherical coordinates associated with the frame $\{X^i\}$.
The triad $\{\bm{N}, \bm{P}, \bm{Q}\}$ is specified by the inclination
$\Theta$ and azimuthal angle $\Phi$ of the unit vector $\bm{N}$. Similarly,
the mass-point location is determined by the pair of angles
$(\theta=\pi/2,\varphi)$, where $\varphi$ is the orbital phase. The
counterparts of those quantities for the reversed observer are $\Theta'=\pi-
\Theta$, $\Phi'= \Phi+ \pi$, $\theta'=\pi/2$ and $\varphi'= \varphi +\pi$.
Note that the components of the relative position $\bm{x}$ and velocity
$\bm{v}$ are solely functions of the azimuthal phase $\varphi$ on the
trajectory $r=r(\varphi)$. Now as is clear from the geometry of the planar
binary, the invariance under rotation implies that
$h(R,T,\Theta,\Phi,\varphi)$ depends on $\Phi$ and $\varphi$ only through
their difference $\varphi-\Phi$. Hence we define as a convenient phase
variable $\phi\equiv\varphi-\Phi+\pi/2$, which is actually the orbital phase
whose origin is chosen to be the direction of the ascending node $\bm{P}$ (see
Fig.~\ref{fig:frame}).
%$h(R,T,\Theta,\Phi,\varphi)=h(R,T,\Theta,0,\varphi-\Phi)$. 
We shall regard $h$ as a function of $\Theta$ and $\phi$, denoted by
$h(\Theta,\phi)$ hitherto, ignoring dependences in $R$, $T$ since they do not
play a role here. Putting those results together, we arrive at the important
identity
\begin{equation}\label{eq:id}
h(\pi-\Theta, \phi) = \overline{h}(\Theta, \phi) \, .
\end{equation}
Inserting the definition \eqref{eq:hlm} of $h^{\ell m}$ into both sides of the
latter equation and noticing~\cite{GMNRS67} that $\overline{Y}^{\ell
  m}_{-2}(\pi-\Theta,\Phi+\pi) = (-1)^{m+\ell}
Y^{\ell\,-m}_{-2}(\Theta,\Phi)$, we conclude from the uniqueness of the
decomposition in spin-weighted spherical harmonics that
\begin{equation}\label{eq:parity_hlm0} 
h^{\ell m}(\varphi) = (-1)^{\ell+m} \,\overline{h}^{\ell,-m}(\varphi+\pi) \,.
\end{equation}
The known dependence of the mode on the azimuthal angle, $h^{\ell
  m}(\varphi)\propto e^{-\ui m \varphi}$, then shows that
\begin{equation}\label{eq:parity_hlm} 
h^{\ell m}(\varphi) = (-1)^{\ell}
  \,\overline{h}^{\ell,-m}(\varphi) \, .
\end{equation}
In the end, we substitute for $h^{\ell m}$ its expression \eqref{eq:inv} in
the equality above. The tensor $\alpha_L^{\ell m}$ introduced in
Eqs.~\eqref{eq:UV} satisfies $\overline{\alpha}_L^{\ell,-m} = (-1)^m
\,\alpha_L^{\ell m}$,\footnote{See Eq.~(2.12) in \cite{Th80}. The notation
  used in Refs.~\cite{Th80,K08} is related to ours by $\mathcal{Y}_L^{\ell
    m}=\frac{(2\ell+1)!!}{4\pi\ell!}\,\overline{\alpha}_L^{\ell m}$.} which
implies that the multipolar coefficients $U^{\ell m}$ and $V^{\ell m}$
satisfy themselves the relations $\overline{U}^{\ell,-m}= (-1)^m\,U^{\ell m}$ and
$\overline{V}^{\ell,-m}= (-1)^m \,V^{\ell m}$. Finally, we find that these
multipole coefficients must be such that $U^{\ell m}= (-1)^{\ell+m} \,U^{\ell m}$
and $V^{\ell m}= (-1)^{\ell+m+1} \,V^{\ell m}$, hence we have necessarily $U^{\ell m}=0$
when $\ell + m$ is odd, and $V^{\ell m} = 0$ when $\ell + m$ is even. We
therefore conclude that there is no mixture between the mass-type and
current-type contributions to the modes, in the sense that
\begin{subequations}\label{eq:hlmUV}\begin{align}
h^{\ell m} &= - \frac{G}{\sqrt{2} R c^{\ell +2}} \,U^{\ell m} \qquad\text{when
  $\ell+m$ is even} \,,\\
h^{\ell m} &= \frac{G}{\sqrt{2} R c^{\ell +3}} \,\ui\,V^{\ell m}
~\qquad\text{when $\ell+m$ is odd} \, .
\end{align} 
\end{subequations}
Let us stress again that this result holds only when the orbit stays planar
and, in particular, does not apply to precessing binaries. However, it is
valid for quasi-circular as well as elliptical orbits.

\section{Expression of the dominant quadrupole mode}\label{sec:quadmode}

In the following sections, we shall compute the dominant quadrupole mode
$(\ell,m)=(2,2)$ at 3.5PN order. From the results \eqref{eq:hlmUV}, the mode
$h^{22}$ depends only on the radiative mass-type quadrupole moment
$U_{ij}$,\footnote{The STF tensor $\alpha^{22}_{ij}$ therein is explicitly
  given by
$$\alpha^{22}_{ij} = \sqrt{\frac{2\pi}{15}}\Bigl(\delta_{\langle
  i}^1-\ui\delta_{\langle
  i}^2\Bigr)\Bigl(\delta_{j\rangle}^1-\ui\delta_{j\rangle}^2\Bigr)\,,$$
with the brackets surrounding indices referring to the STF projection.}
\begin{equation}\label{eq:h22}
h^{22} = -\frac{G}{\sqrt{2}R c^4} \,U^{22} = - \frac{\sqrt{6}G}{R c^4}
\,\alpha^{22}_{ij}\,U_{ij} \,.
\end{equation}
Following the multipolar-post-Minkowskian (MPM)
formalism~\cite{BD86,BD92,B98quad,B98tail,B98mult}, we shall express the
radiative quadrupole moment $U_{ij}$ in terms successively of canonical
moments $\{M_L, S_L\}$ and then source moments $\{I_L, J_L, W_L, X_L, Y_L,
Z_L\}$ for general sources at 3.5PN order. Among the source moments, $\{I_L,
J_L\}$ are the most important ones because they arise at Newtonian order,
while $\{W_L, X_L, Y_L, Z_L\}$, which are called the gauge moments, induce
corrections starting at high 2.5PN order.

Using the ``selection rules'' developed in Paper I, we can identify all types
of terms to be computed up to quadratic order (which is sufficient for our
purpose). Let $\ell_\text{max} [A_J,B_K]$ be the maximum multipolar order
beyond which $U_L$ cannot contain products of the two multipole moments $A_J$
and $B_K$ or their time derivatives (or possibly time anti-derivatives), at
the 3.5PN level. Here $A_J$ and $B_K$ can be any moments chosen among the set of
source moments $\{I_L, J_L, W_L, X_L, Y_L, Z_L\}$. Thus when
$\ell>\ell_\text{max} [A_J,B_K]$ any term in $U_L$ constructed with the
product of moments $A_J$ and $B_K$ has a post-Newtonian order strictly higher
than 3.5PN and is negligible here. On the other hand when $\ell_\text{max}
[A_J,B_K]<2$,
the corresponding interaction is not allowed to enter the quadrupole moment
$U_{ij}$ which has $\ell=2$. When $\ell_\text{max} [A_J,B_K]\geqslant 2$,
it is equal to the number of (free or dummy) indices to be shared between: (i)
$A_J$ and $B_K$ if both moments are of the same parity, \textit{i.e.}
$j+k=\ell_\text{max}[A_J,B_K]$, or (ii) $A_{aJ-1}$, $B_{bK-1}$ and
$\varepsilon_{iab}$ (with contractions on indices $a$ and $b$) if both moments
are of different parity, \textit{i.e.}
$j+k=\ell_\text{max}[A_J,B_K]+1$. The values of $\ell_\text{max}[A_J,B_K]$
larger than or equal to 2
are listed in Table~\ref{tab:lmax} for the various possible choices of $A_J$
and $B_K$. Take for instance the case of $I_J$ and $I_K$. By dimensionality we
can add the powers of $1/c$ and find that this term is necessarily of the type
($p$ and $q$ being the number of time derivatives)
\begin{equation}\label{eq:terme}
U_L \sim \frac{G}{c^{j+k+3-\ell}}\,I_J^{(p)} I_K^{(q)}\,.
\end{equation}
The contribution of this term in the waveform \eqref{eq:hij} is easily seen to
be of post-Newtonian order $\mathcal{O}(1/c^{j+k+1})$ and hence we require
$j+k=6$ for a 3.5PN term. On the other hand, we have said that
$\ell_\text{max}=j+k$ in this case, hence we have $\ell_\text{max}=6$ in
Table~\ref{tab:lmax} for a contribution like \eqref{eq:terme}. Similar
selection rules can be made for the radiative current moment $V_L$ but here we
need only to compute the mass quadrupole moment. From Table~\ref{tab:lmax}, it
is straightforward to obtain the set of quadratic interactions contributing to
$U_{ij}$ within our accuracy. No new cubic interaction (between three source
moments), besides the already known 3PN tail-of-tail term, enters the mass
quadrupole $U_{ij}$ at the 3.5PN order.
\begin{table}[here]
\begin{tabular}{|l||c|c|c|c|c|c|} 
\hline
$A_J$ & \multicolumn{6}{c|}{$B_K$} \\ \cline{2-7} %\hline \hline 
      &$I_K$&$J_K$&$W_K$&$X_K$&$Y_K$&$Z_K$ \\ \hline 
$I_J$ &  6  &  4  &  4  &  2 &  4  &  2  \\ \hline
$J_J$ &  4  &  4  &  2  & -- &  2  &  2  \\ \hline
$W_J$ &  4  &  2  &  2  & -- &  2  & --  \\ \hline
$X_J$ &  2  & --  &  -- & -- & --  & --  \\ \hline
$Y_J$ &  4  &  2  &  2  & -- &  2  & --  \\ \hline
$Z_J$ &  2  &  2  &  -- & -- & --  & --  \\ \hline
\end{tabular}
\caption{Values of $\ell_\text{max}[A_J,B_K]$ at the 3.5PN order for the
  various possible choices of $A_J$ and $B_K$. Because factors may always be
  reordered in products, we have $\ell_\text{max}[A_J,B_K] =
  \ell_\text{max}[B_J,A_K]$ and the table is trivially
  symmetric.} \label{tab:lmax}
\end{table}

The long calculation following the MPM algorithm~\cite{BD86} (in the slightly
modified version of Ref.~\cite{B98quad}), is implemented in a systematic way
by means of the tensor package xTensor developed for Mathematica
\cite{xtensor}, and yields various types of terms:
\begin{equation}\label{eq:U2decomp}
U_{ij} = U^\text{inst}_{ij} + U^\text{tail}_{ij} + U^\text{mem}_{ij}\,.
\end{equation}
In a first stage we express each of the different pieces by means of the
canonical moments $\{M_L, S_L\}$. The instantaneous piece up to 3.5PN order is
\begin{align}\label{eq:U2inst}
U_{ij}^\text{inst} &= M^{(2)}_{ij} \nonumber \\ &+\frac{G}{
c^5}\biggl[ \frac{1}{
7}M^{(5)}_{a\langle i}M_{j\rangle a} - \frac{5}{7} M^{(4)}_{a\langle
i}M^{(1)}_{j\rangle a} -\frac{2}{7} M^{(3)}_{a\langle i}M^{(2)}_{j\rangle a}
+\frac{1}{3}\varepsilon_{ab\langle i}M^{(4)}_{j\rangle a}S_{b}\biggr]\nonumber
\\ 
&+ \frac{G}{c^7} \bigg[ -
\frac{64}{63} S^{(2)}_{a{\langle i}}   
 S^{(3)}_{{j\rangle}a}  + \frac{1957}{3024}  M^{(3)}_{ijab}  
  M^{(4)}_{ab}  + \frac{5}{2268} 
  M^{(3)}_{ab{\langle i}}    M^{(4)}_{{j\rangle}ab}  + 
\frac{19}{648}  M^{(3)}_{ab}  
 M^{(4)}_{ijab}  \nonumber\\ & \qquad\quad + \frac{16}{63} 
 S^{(1)}_{a{\langle i}}    S^{(4)}_{{j\rangle}a} + \frac{1685}{1008} M^{(2)}_{ijab}  
 M^{(5)}_{ab}  + \frac{5}{126}  M^{(2)}_{ab{\langle i}}  
 M^{(5)}_{{j\rangle}ab} -\frac{5}{756} 
 M^{(2)}_{ab}   M^{(5)}_{ijab}  \nonumber\\ & \qquad\quad + \frac{80}{63}
S_{a{\langle i}}  S^{(5)}_{{j\rangle}a} +\frac{5}{42} S_a 
 S^{(5)}_{ija}  + \frac{41}{28}  M^{(1)}_{ijab}  
  M^{(6)}_{ab}   + \frac{5}{189} 
 M^{(1)}_{ab{\langle i}}    M^{(6)}_{{j\rangle}ab}  \nonumber\\ & \qquad\quad +
\frac{1}{432} M^{(1)}_{ab}  
 M^{(6)}_{ijab}   + \frac{91}{216} M_{ijab}  
 M^{(7)}_{ab}  - \frac{5}{252} M_{ab{\langle i}} 
 M^{(7)}_{{j\rangle}ab}  - \frac{1}{432}
M_{ab}  M^{(7)}_{ijab} \nonumber\\ & \qquad\quad + \varepsilon_{ac{\langle i}}
\Big(\frac{32}{189}  
   M^{(3)}_{{j\rangle}bc}   S^{(3)}_{ab}   -
  \frac{1}{6}  M^{(3)}_{ab}  
 S^{(3)}_{{j\rangle}bc} + \frac{3}{56}   S^{(2)}_{{j\rangle}bc}  
 M^{(4)}_{ab}  + \frac{10}{189}   S^{(2)}_{ab}  
 M^{(4)}_{{j\rangle}bc}  \nonumber\\ & \qquad\quad + \frac{65}{189}  
 M^{(2)}_{{j\rangle}bc}    S^{(4)}_{ab}  + 
\frac{1}{28}   M^{(2)}_{ab}   S^{(4)}_{{j\rangle}bc}  + 
\frac{187}{168}  S^{(1)}_{{j\rangle}bc}   M^{(5)}_{ab} 
 - \frac{1}{189}   S^{(1)}_{ab}   
 M^{(5)}_{{j\rangle}bc}  \nonumber\\ & \qquad\quad - \frac{5}{189}  M^{(1)}_{{j\rangle}bc}  
 S^{(5)}_{ab}  +\frac{1}{24}   M^{(1)}_{ab}  
  S^{(5)}_{{j\rangle}bc}  + \frac{65}{84} S_{{j\rangle}bc} 
  M^{(6)}_{ab}  + \frac{1}{189} S_{ab}  
 M^{(6)}_{{j\rangle}bc} \nonumber\\ & \qquad\quad -\frac{10}{63} M_{{j\rangle}bc}  
 S^{(6)}_{ab} +\frac{1}{168} M_{ab} 
 S^{(6)}_{{j\rangle}bc} \Big) \bigg] \,,
\end{align}
where the angle brackets $\langle\, \rangle$ around indices $i$ and $j$ denote
their STF projection (for convenience we do not indicate the neglected
remainder $\mathcal{O}(1/c^8)$). The hereditary tail integrals involve the
dominant tail term at 1.5PN order \cite{BD92} and the tail-of-tail term at 3PN
order \cite{B98tail}:
\begin{align}\label{eq:U2tail}
U^\text{tail}_{ij} &= \frac{2G M}{c^3} \int_{-\infty}^{T_R} \ud
\tau \left[ \ln \left(\frac{T_R-\tau}{2\tau_0}\right)+\frac{11}{12} \right]
M^{(4)}_{ij} (\tau) \nonumber \\ 
&+ 2\left(\frac{G M}{c^3}\right)^2\int_{-\infty}^{T_R} \ud \tau \left[ \ln^2
\left(\frac{T_R-\tau}{2\tau_0}\right)+\frac{57}{70} \ln\left(\frac{T_R-\tau}{
2\tau_0}\right)+\frac{124627}{44100} \right] M^{(5)}_{ij} (\tau)\,,
\end{align}
where $\tau_0$ denotes an arbitrary constant which will enter \textit{in fine}
into some unphysical frequency scale \eqref{eq:omega0} in the waveform. The
memory-type hereditary integrals, which constitute the third piece in
Eq.~\eqref{eq:U2decomp}, contribute here at orders 2.5PN and 3.5PN. Our
expression extends the classic works on the dominant 2.5PN non-linear memory
effect~\cite{BD92,Chr91,Th92,WW91,B98quad}, and is in agreement with the
recent computation of the non-linear memory up to any post-Newtonian order in
Refs.~\cite{F09,F11}. We have
\begin{align}\label{eq:U2mem}
  U_{ij}^\text{mem} &= \frac{G}{c^5}\biggl[-\frac{2}{7}\int_{-\infty}^{T_R}
  \ud\tau\, M^{(3)}_{a\langle
    i}(\tau)\,M^{(3)}_{j\rangle a}(\tau) \biggr]\nonumber \\
  &+ \frac{G}{c^7} \bigg[-\frac{32}{63} \int_{-\infty }^{T_R} \ud\tau\,
  S^{(3)}_{a{\langle i}}(\tau) \,S^{(3)}_{{j \rangle}a}(\tau) - \frac{5}{756}
  \int_{-\infty }^{T_R} \ud\tau\, M^{(4)}_{ab}(\tau) \,M^{(4)}_{ijab}(\tau)
  \nonumber\\ & \qquad\quad - \frac{20}{189} \,\varepsilon_{ab{\langle i}}
  \int_{-\infty}^{T_R} \ud\tau\, S^{(3)}_{ac}(\tau) \,M^{(4)}_{{j
      \rangle}bc}(\tau) + \frac{5}{42} \,\varepsilon_{ab{\langle i}}
  \int_{-\infty }^{T_R} \ud\tau \,M^{(3)}_{ac}(\tau)
  \,S^{(4)}_{{j \rangle}bc}(\tau) \biggr] \,.
\end{align}
In a second stage of the general formalism, we must express the canonical
moments $\{M_L, S_L\}$ in terms of the six types of source moments $\{I_L,
J_L, W_L, X_L, Y_L, Z_L\}$. For the present computation we need only to relate
the canonical quadrupole moment $M_{ij}$ to the corresponding source
quadrupole moment $I_{ij}$ up to 3.5PN order. We obtain
\begin{align}\label{eq:cansource}
  M_{ij} &= I_{ij} +\frac{4G}{c^5}
  \left[W^{(2)}I_{ij}-W^{(1)}I_{ij}^{(1)}\right] \nonumber\\ 
  &+ \frac{4G}{c^7} \biggl[ \frac{4}{7} W^{(1)}_{a{\langle i}} I^{(3)}_{{j
      \rangle}a} + \frac{6}{7} W_{a{\langle i}} I^{(4)}_{{j \rangle}a} -
  \frac{1}{7} I_{a{\langle i}} Y^{(3)}_{{j \rangle}a} - Y_{a{\langle i}}
  I^{(3)}_{{j \rangle}a} - 2 X I^{(3)}_{ij} - \frac{5}{21} I_{ija} W^{(4)}_a \nonumber\\ &
  \qquad\quad + \frac{1}{63} I^{(1)}_{ija} W^{(3)}_a - \frac{25}{21} I_{ija}
  Y^{(3)}_a - \frac{22}{63} I^{(1)}_{ija} Y^{(2)}_a + \frac{5}{63} Y^{(1)}_a
  I^{(2)}_{ija} + 2  W_{ij} W^{(3)} \nonumber \\ & \qquad\quad +
2 W_{ij}^{(1)} W^{(2)} - \frac{4}{3} W_{{\langle i}} W_{{j \rangle}}^{(3)} + 2 Y_{ij} W^{(2)} -
4 W_{\langle i} Y^{(2)}_{j \rangle}
 \nonumber\\ & \qquad\quad + \varepsilon_{ab{\langle i}}
  \biggl(\frac{1}{3} I_{{j \rangle}a} Z^{(3)}_b + \frac{4}{9} J_{{j \rangle}a}
  W^{(3)}_b - \frac{4}{9} J_{{j \rangle}a} Y^{(2)}_b 
+ \frac{8}{9} J^{(1)}_{{j \rangle}a} Y^{(1)}_b + Z_a I^{(3)}_{{j \rangle}b} \biggr)
  \biggr]\,.
\end{align}
The current quadrupole canonical moment $S_{ij}$ and mass octupole canonical
moment $M_{ijk}$ agree with the corresponding source moments up to 2PN order,
and for this computation it is sufficient to replace these by their source
counterparts.
 
\section{Source multipole moments of compact
  binaries}\label{sec:sourcemoments}

We now feed the previous expression of the radiative quadrupole moment
$U_{ij}$ with explicit results for the source moments in the case of
non-spinning compact binaries on quasi-circular orbits. The general formulas
for the six sets of source moments $\{I_L, J_L, W_L, X_L, Y_L, Z_L\}$ as
explicit integrals over the matter distribution in the source and the
gravitational field were obtained in Ref.~\cite{B98mult}. The mass-type source
moments expressed in terms of elementary retarded potentials up to 3.5PN order
are given by Eqs.~(3.3)--(3.6) of Ref.~\cite{BI04mult}. Reducing the latter
source quadrupole moment $I_{ij}$ for compact binaries, in the frame of the
center-of-mass and for quasi-circular orbits, yields
\begin{equation}\label{eq:I3PN}
I_{ij} = \nu\,m\,\left(\mathcal{A}\,x^{\langle
i}x^{j\rangle} +\frac{r^2}{c^2} \,\mathcal{B}\,v^{\langle
i}v^{j\rangle} + \frac{48}{7}\frac{G^2 m^2 \nu}{c^5 r}
\,\mathcal{C}_\text{RR}\, x^{\langle i}v^{j\rangle}\right)+
\mathcal{O}\left(\frac{1}{c^8}\right)\,,
\end{equation} 
where $x^i$ is the orbital separation and $v^i=\ud x^i/\ud t$ is the relative
velocity. Resorting also to the
post-Newtonian parameter $\gamma$ defined by Eq.~\eqref{eq:PNgam}, the
``conservative'' coefficients $\mathcal{A}$ and $\mathcal{B}$ in
\eqref{eq:I3PN} read
\begin{subequations}\label{eq:AB}\begin{align}
\mathcal{A} &= 1 + \gamma \left(-\frac{1}{ 42}-
\frac{13}{ 14}\nu \right) + \gamma^2
    \left(-\frac{461}{ 1512} -\frac{18395}{ 1512}\nu - \frac{241}{ 1512}
      \nu^2\right) \nonumber \\ &~~+ \gamma^3 \left(\frac{395899}{
        13200}-\frac{428}{ 105}\ln \left(\frac{r}{r_0}\right)
      +\left[\frac{3304319}{ 166320} - \frac{44}{ 3}\ln \left(\frac{r}{
            r'_0}\right)\right]\nu \right.\nonumber\\ &~~\qquad + \left.
      \frac{162539}{ 16632} \nu^2 + \frac{2351}{ 33264}\nu^3 \right) \,,\\ 
\mathcal{B}
    &= \frac{11}{ 21}-\frac{11}{ 7}\nu  + \gamma
    \left(\frac{1607}{ 378}-\frac{1681}{ 378} \nu +\frac{229}{ 378}\nu^2\right)
    \nonumber \\ &~~+ \gamma^2 \left(-\frac{357761}{ 19800}+\frac{428}{ 105}
      \ln \left(\frac{r}{r_0} \right) - \frac{92339}{ 5544} \nu + \frac{35759}{
        924} \nu^2 + \frac{457}{ 5544} \nu^3 \right)\,.
\end{align}\end{subequations}
The results at 3PN order were already known \cite{BIJ02,BI04mult}. Note the
appearance at 3PN order of two arbitrary scales: the first one is given by
$r_0=c\tau_0$, where $\tau_0$ is the same scale as in the tail integrals
\eqref{eq:U2tail}; the second one is a Hadamard regularization scale $r'_0$
due to the presence of the point mass singularities. The scale $r_0$ is
associated with an irrelevant choice of the origin of time in the radiative
zone, whereas the scale $r'_0$ will disappear from our final results, being
cancelled by the corresponding $r'_0$ in the equations of motion; see
Eq.~\eqref{eq:omega3PN}. The new correction term at 3.5PN order only affects,
for circular orbits, the ``radiation reaction'' coefficient
$\mathcal{C}_\text{RR}$ which is found to be
\begin{equation}\label{eq:C}
\mathcal{C}_\text{RR} = 1+\gamma\left(
        -\frac{256}{135}-\frac{1532}{405}\nu \right) \,.
\end{equation}
We give the result for the complete 3.5PN mass quadrupole moment in the case
of generic orbits reduced to the center-of-mass frame in
Appendix~\ref{sec:appA}.

In addition to the 3.5PN mass quadrupole $I_{ij}$ we will also need the
expression of the mass monopole or ADM mass $M$ (notice that $I\equiv M$) up
to 2PN order since it is required in the computation of the tail terms
\eqref{eq:U2tail}. We have, for circular orbits:
\begin{equation}\label{eq:M2PN}
  M = m
  \biggl[1-\frac{\nu}{2}\gamma+\frac{\nu}{8}\bigl(7-\nu\bigr)\gamma^2\biggr] +
  \mathcal{O}\left(\frac{1}{c^5}\right)\,.
\end{equation}
Furthermore the current dipole moment $J_i$ is required up to 1PN order, since
it appears in the 2.5PN term of Eq.~\eqref{eq:U2inst} (recall that $S_i=J_i$
in our approximation):
\begin{equation}\label{eq:Ji1PN}
  J_i = \nu\,m \,\varepsilon_{ijk}\,x^jv^k
  \left[1+\gamma\left(\frac{7}{2}-\frac{1}{2}\nu\right)\right] +
  \mathcal{O}\left(\frac{1}{c^4}\right) \, .
\end{equation}
The other source moments $I_L$ and $J_L$ are required at Newtonian order only,
and at that order we have the following general formulas, valid for all
multipolar indices $\ell\geqslant 2$:
\begin{subequations}\label{eq:sourcemom}
\begin{align}
  I_L &= \nu\,m\,s_\ell(\nu)\,\hat{x}_L +
  \mathcal{O}\left(\frac{1}{c^2}\right)\,, \\
  J_L &= \nu\,m\,s_{\ell+1}(\nu)\,L_{\langle i_\ell}\hat{x}_{L-1\rangle}+
  \mathcal{O}\left(\frac{1}{c^2}\right)\,.
\end{align}\end{subequations}
Here we denote $L_i = \varepsilon_{ijk}\, x^j v^k$ and $s_{\ell}(\nu) =
X_2^{\ell-1}+(-1)^\ell X_1^{\ell-1}$, with
$X_1=\frac{m_1}{m}=\frac{1}{2}(1+\Delta)$ and
$X_2=\frac{m_2}{m}=\frac{1}{2}(1-\Delta)$ where
$\Delta=\frac{m_1-m_2}{m}=\pm\sqrt{1-4\nu}$; see also Footnote $24$ of Paper I
for an alternative expression of the function $s_{\ell}(\nu)$.

Among the required gauge-type source moments $\{W_L, X_L, Y_L, Z_L\}$, the only
one which is to be controlled at 1PN order is the monopole $W$, since it
enters the 2.5PN term of Eq.~\eqref{eq:cansource}. However it turns out to be
zero for quasi-circular orbits. The other needed moments are merely Newtonian.
Their full list is as follows:
\begin{subequations}\label{eq:gaugemom}
\begin{align}
W &= 0 + \mathcal{O}\left(\frac{1}{c^4}\right) \,,\\
W_{i} &= \frac{1}{10} m\,\nu\,\Delta\,r^2\,v^i
  +\mathcal{O}\left(\frac{1}{c^2}\right) \,,\\
W_{ij} &= - \frac{2}{21} m\,\nu\bigl(1-3\nu\bigr) r^2 x^{\langle
    i}v^{j\rangle} +\mathcal{O}\left(\frac{1}{c^2}\right) \,,\\ 
X &=\frac{1}{12} G m^2 r \,\nu \,\bigl(-1 + \nu\bigr)  +
\mathcal{O}\left(\frac{1}{c^2}\right) \,,\\
Y_{i} &= \frac{G m^2\nu}{5 r}\Delta\,x^i +
\mathcal{O}\left(\frac{1}{c^2}\right) \,,\\   
Y_{ij} &= \frac{1}{14}\frac{G m^2\nu}{r}\bigl(-3+16\nu\bigr) x^{\langle
    i}x^{j\rangle} - \frac{2}{7} m\,\nu\bigl(1-3\nu\bigr)r^2\,v^{\langle
    i}v^{j\rangle} +\mathcal{O}\left(\frac{1}{c^2}\right) \,,\\
Z_i &= 0 + \mathcal{O}\left(\frac{1}{c^2}\right)\,.
\end{align}\end{subequations}

Finally, when computing the waveform, we need the multiple time derivatives of
all these moments; they are computed by order reduction using the acceleration
for quasi-circular orbits given at 3.5PN order by Eq.~\eqref{eq:acircfinal}.
Furthermore, we need to perform many scalar products between the velocity and
the polarization vectors $(\bm{N}, \bm{P}, \bm{Q})$; these are computed using
the 3.5PN expression of the velocity as given by Eq.~\eqref{eq:vcircfinal}.
Note also that the (tails and non-linear memory) hereditary terms given by
Eqs.~\eqref{eq:U2tail} and \eqref{eq:U2mem} are treated by exactly the same
method as in Paper I.

\section{Dominant quadrupolar mode at 3.5PN order}\label{sec:h22}

We shall now perform a change of phase variable, from the actual orbital phase
related to the orbital frequency by $\varphi=\int\omega\ud t$, to a new phase
variable $\psi$, defined in such a way that it absorbs most of the logarithmic
contributions in the frequency coming from the tail terms. Extending
Refs.~\cite{BIWW96,ABIQ04,BFIS08}, we find that the new phase variable takes
remarkably the same expression as the one used in Paper I:
\begin{equation}\label{eq:changephaseomega}
\psi = \varphi - \frac{2 G M \omega}{c^3}\ln\left(\frac{\omega}{\omega_0}\right)
\, .
\end{equation}
The 2PN corrections in the ADM mass $M$, as given by Eq.~\eqref{eq:M2PN}, are
exactly the ones needed for the logarithmic cancellations at our extended
3.5PN order. Here $\omega_0$ is a constant frequency scale related to the
constant $\tau_0$ appearing in the tail integrals \eqref{eq:U2tail} by
\begin{equation}\label{eq:omega0}
\omega_0 = \frac{e^{\frac{11}{12}-\gamma_\text{E}}}{4\tau_0} \,,
\end{equation}
with $\gamma_\text{E}=0.577\cdots$ denoting the Euler constant. It is
known~\cite{BIWW96} that this frequency scale is irrelevant: a rescaling of
$\omega_0$ is indeed equivalent to a shift of the binary's instant of
coalescence, which can be absorbed into a redefinition of the origin of time
in the radiation zone. Physically, the logarithmic term in
Eq.~\eqref{eq:changephaseomega} expresses the fact that the tails induce a
small delay in the arrival time of gravitational waves. When viewed as a
modulation of the orbital phase as in \eqref{eq:changephaseomega}, this term
represents in fact a very small correction to this phase, being of order 4PN
when compared to the dominant phase evolution which is at the ``inverse'' of
2.5PN order (\textit{i.e.} at order $c^{+5}$), and is negligible with the
present accuracy.

Replacing $\varphi$ in terms of the new phase variable $\psi$, we perform the
appropriate Taylor expansion of the waveform and apply the mode decomposition
\eqref{eq:hdecomp}. We also use the fact that, from the planar geometry of the
problem, $h=h_+-\ui h_\times$ depends only on the
inclination angle $\Theta$ and the difference $\psi-\Phi$ between the
``absolute'' phase $\psi$ and the azimuth $\Phi$ of the vector $\bm{N}$ (see
also Sec.~\ref{sec:gwf} and Fig.~\ref{fig:frame}). For instance, like in
Eq.~\eqref{eq:id} but with $\psi$ playing now the role of $\varphi$, we can
introduce the phase $\tilde{\phi}=\psi-\Phi+\pi/2$ defined relatively to the
direction of the ascending node $\bm{P}$. Choosing the latter phase as
integration variable instead of $\Phi$, and using the known dependence of the
spherical harmonics on the angle $\tilde{\phi}$, we obtain
\begin{equation}\label{eq:hlmmode}
h^{\ell m}(\psi) = (-\ui)^m\,e^{-\ui m \,\psi}\int \ud\Omega
\,h(\Theta,\tilde{\phi})\,Y^{\,\ell m}_{-2} (\Theta,\tilde{\phi})\,,
\end{equation}
exhibiting the azimuthal factor $e^{-\ui m \,\psi}$ appropriate for each mode.
Alternatively, we may access directly a given $h^{\ell m}$ mode, once the
requisite radiative multipole $U_L$ or $V_L$ known, by using the contraction
formulas \eqref{eq:UV}, which translate into \eqref{eq:h22} for $h^{22}$.
Finally, to present our result we pose
\begin{equation}\label{eq:modedef}
  h^{\ell m} = \frac{2 G \,m \,\nu \,x}{R \,c^2} \,\sqrt{\frac{16\pi}{5}}\,
H^{\ell m}\,e^{-\ui m \, \psi} \,,
\end{equation}
and express it entirely with the help of the gauge invariant post-Newtonian
parameter $x$ related to the orbital frequency by Eq.~\eqref{eq:PNx}. We find
for the dominant $(2,2)$ mode up to 3.5PN order (extending the 3PN result of
Paper I for this mode):
\begin{align} \label{eq:h22res}
H^{22} &= 1+x \left(-\frac{107}{42}+\frac{55}{42}\nu\right)+2 \pi
  x^{3/2}+x^2
  \left(-\frac{2173}{1512}-\frac{1069}{216}\nu+\frac{2047}{1512}\nu^2\right)
  \nonumber \\ 
  &+ x^{5/2} \left(-\frac{107 \pi }{21}-24 \,\ui\,\nu +\frac{34
      \pi}{21}\nu\right)+x^3
  \bigg(\frac{27027409}{646800}-\frac{856}{105}\,\gamma_\text{E} +\frac{428 
    \,\ui\,\pi }{105}+\frac{2 \pi ^2}{3}\nonumber \\
  &\qquad+ \left(-\frac{278185}{33264}+\frac{41 \pi^2}{96}\right) \nu
  -\frac{20261}{2772}\nu^2+\frac{114635}{99792}\nu^3-\frac{428}{105} \ln (16
  x)\bigg) 
  \nonumber \\
  &+ x^{7/2} \left( -\frac{2173\pi}{756} + \left(
      -\frac{2495\pi}{378}+\frac{14333\,\ui}{162} \right)\nu + \left(
      \frac{40\pi}{27}-\frac{4066\,\ui}{945} \right)\nu^2 \right) \,.
\end{align}
We have verified that the above expression is in full agreement in the test mass
limit where $\nu\to 0$, with the result of black-hole perturbation theory as
reported in the Appendix B of Ref.~\cite{TSasa94} or, in a form more suitable
for comparison, in Eq.~(4.9a) of Ref.~\cite{FI10}.

As it is the dominant quadrupole $(2,2)$ mode that is determined with the best
precision by numerical relativity, the 3.5PN result \eqref{eq:h22res} should
become important for high-accuracy comparisons between numerical and
analytical waveforms. See Paper I for all the other modes that contribute up
to 3PN order.
%\footnote{We have recently spotted an error in the 2.5PN expression of the
%current quadrupole source moment $J_{ij}$ computed in Paper I. As a result
%one of the coefficients in the $(2,1)$ mode should be corrected: namely
%$-995/84$ should read $-321/28$ in Eq.~(9.4b) of Paper I; see the published
%Erratum~\cite{BFIS08}.} 
The completion of the remaining modes at 3.5PN order presents some new
difficulties; in particular the $(2,1)$ mode necessitates the
difficult extension of the current quadrupole moment $J_{ij}$ up to 3PN order.
This is left for future work.

\acknowledgments 
GF, LB, and BRI thank the Indo-French Collaboration (IFCPAR) under which
this work has been carried out.

\appendix

\section{3.5PN mass quadrupole moment}\label{sec:appA}

We give our result for the complete source mass quadrupole $I_{ij}$ up to
3.5PN order, for general non-circular orbits, but reduced to the center-of-mass
frame. Defining:
\begin{align}\label{eq:Iijdefinitions}
I_{ij} &= \nu\,m\,
 \left[ \left(\mathcal{A}-\frac{24}{7}\frac{G^2
m^2\nu}{c^5 r^2}\,\dot{r}\mathcal{A}_\text{RR}\right)\, x^{\langle
i}x^{j\rangle} + \left( \frac{r^2}{c^2}\,\mathcal{B} + \frac{G
m \nu}{c^5}\,r\,\dot{r}\,\mathcal{B}_\text{RR} \right) v^{\langle i}v^{j\rangle} \right.
\nonumber\\
&  \left. \qquad\quad +
2\left(\frac{r \dot{r}}{c^2}\,\mathcal{C}+\frac{24}{7}
 \frac{G^2 m^2\nu}{c^5 r}\mathcal{C}_\text{RR} \right) x^{\langle i}v^{j\rangle}
\right] \, ,
\end{align}
the expressions for the conservative parts $\mathcal{A}$, $\mathcal{B}$,
and $\mathcal{C}$ read:
\begin{subequations}\label{eq:ABCcons}\begin{align}
%%---------------------------0PN term--------------------------
\mathcal{A}&= 1
%----------------------------1PNterms below-------------
+\frac{1}{c^2} \left[v^2\, \left( \frac{29}{42}
  - \frac{29\,\nu }{14} \right)+\frac{G\,m}{r}\, \left( -\frac{5}{7}+
  \frac{8 }{7}\,\nu \ \right) \right]\nonumber\\
%----------------------------2PNterms below-------------
&+ \frac{1}{c^4}\left[v^2\,\frac{G\,m}{r}\,\left( \frac{2021}{756} -
\frac{5947 }{756}\,\nu-\frac{4883}{756}\,\nu^2
\right)\right.\nonumber\\
&\qquad\left.+\left(\frac{G\,m}{r}\right)^2\,\left(
- \frac{355}{252} -\frac{953 }{126}\,\nu + \frac{337\,}{252}\,\nu^2
\right)\right.\nonumber\\
&\qquad+\left.v^4\,\left( \frac{253}{504} -
\frac{1835 }{504}\,\nu
+\frac{3545}{504}\,\nu^2\right)\right.\nonumber\\
&\qquad
\left. +\dot{r}^2\, \frac{G\,m}{r}\,\left( - \frac{131}{756} +
\frac{907 }{756}\,\nu - \frac{1273}{756}\,\nu^2
\right)\right]\nonumber\\
%-----------------------------3PN terms below----------
&+\frac{1}{c^6}\left[v^6\,\left( \frac{4561}{11088} -
\frac{7993}{1584}\,\nu+\frac{117067}{5544}\,\nu^2 -
\frac{328663}{11088}\,\nu^3\right)\right.\nonumber\\
%%%%%%%%%%%%%%%%%%%%%%%%%%%%%%%%%%%%%%%
&\left.\qquad+v^4\, \frac{G\,m}{r}\,\left( \frac{307}{77} - \frac{94475
}{4158}\,\nu+\frac{218411}{8316}\,\nu^2 +
\frac{299857}{8316}\,\nu^3\right)\right.\nonumber\\
%%%%%%%%%%%%%%%%%%%%%%%%%%%%%%%%%%%%%%%%%%%%
&\qquad+\left(\frac{G\,m}{r}\right)^3\,\left(
\frac{6285233}{207900}+\frac{15502}{385}\,\nu
-\frac{3632}{693}\,\nu^2 + \frac{13289}{8316}\,\nu^3
\right.\nonumber\\
&\qquad\quad\left. -\frac{428}{105}\,\ln \left(\frac{r}{r_0}\right) -
  \frac{44}{3}\,\,\nu 
\,\ln \left(\frac{r}{r_0'}\right) \right)\nonumber\\
%%%%%%%%%%%%%%%%%%%%%%%%%%%%%%%%%%%%%%%%%%%%%%%%%%%%%%%%%%%
&\qquad+{\dot{r}}^2\,\left(\frac{G\,m}{r}\right)^2\,\left( -
\frac{8539}{20790}+ \frac{52153 }{4158}\,\nu - \frac{4652}{231}\,\nu^2
-\frac{54121}{5544}\,\nu^3 \right) \,\nonumber\\
&\qquad+{\dot{r}}^4\,\frac{G\,m}{r}\,\left( \frac{2}{99} -
\frac{1745}{2772}\,\nu +\frac{16319}{5544}\,\nu^2 -
\frac{311\,}{99}\,\nu^3 \right) \,\nonumber\\
%%%%%%%%%%%%%%%%%%%%%%%%%%%%%%%%%%%%%%%%%%%%%%%%%%%%%%%%%%%%%
&\qquad+v^2\,\left(\frac{G\,m}{r}\right)^2\,\left( \frac{187183}{83160}
-\frac{605419 }{16632}\,\nu + \frac{434909}{16632}\,\nu^2
-\frac{37369}{2772}\,\nu^3 \right)\nonumber\\
%%%%%%%%%%%%%%%%%%%%%%%%%%%%%%%%%%%%%%%%%%%%%%%%%%%%%%%%%%%%%%%%%%%%%
&\qquad+\left.v^2\,\frac{G\,m}{r}\,\,{\dot{r}}^2 \,\left( -
\frac{757}{5544}+\frac{5545 }{8316}\,\nu -
\frac{98311\,}{16632}\,\nu^2 +\frac{153407}{8316}\,\nu^3 \right)
\right]\,, \\
%@@@@@@@@@@@@@@@@@@@@@ B BEGINS @@@@@@@@@@@@@@@@@@@@@@@@@@@@@@@
\mathcal{B} &= \frac{11}{21} - \frac{11}{7}\,\nu\nonumber\\
&+\frac{1}{c^2}\left[\frac{G m}{r}\,\left( \frac{106}{27} -
\frac{335}{189}\,\nu - \frac{985}{189}\,\nu^2
\right)\right. \nonumber\\
&\qquad\left. +\,v^2\,\left(
\frac{41}{126} - \frac{337 }{126}\,\nu + \frac{733}{126}\,\nu^2
\right)+\dot{r}^2\,\left( \frac{5\,}{63} - \frac{25 }{63}\,\nu +
\frac{25}{63}\,\nu^2 \right)\right] \nonumber\\
&+\frac{1}{c^4}\,\left[\,v^4\,\left( \frac{1369}{5544} - \frac{19351
}{5544}\,\nu + \frac{45421}{2772}\,\nu^2 - \frac{139999}{5544}\,\nu^3
\right) \right.\nonumber\\
%-----------------------------------
&\qquad +\left.\left(\frac{G
m}{r}\right)^2\,\left(-\frac{40716}{1925}- \frac{10762}{2079}\,\nu
+\frac{62576}{2079}\,\nu^2 - \frac{24314}{2079}\,\nu^3
\right.\right.\nonumber\\
&\qquad\quad\left.\left.  +\frac{428}{105}\,\ln\left(\frac{r}{r_0}\right)
\right)\right.\nonumber\\
&\qquad+\dot{r}^2\,\frac{G m}{r}\left(\frac{79}{77}-\frac{5807}{1386}\,\nu +
  \frac{515}{1386}\,\nu^2 
+\frac{8245}{693}\,\nu^3 \right)\nonumber\\
&\qquad+ v^2\,\frac{G m}{r}\,\left( \frac{587}{154} - \frac{67933 }{4158}\,\nu+
\frac{25660}{2079}\,\nu^2 +\frac{129781}{4158}\,\nu^3
\right)\nonumber\\
&\qquad+\left. v^2\,\dot{r}^2\,\left(\frac{115\,}{1386}-\frac{1135}{1386}\,\nu
+\frac{1795}{693}\,\nu^2-\frac{3445}{1386}\,\nu^3 \right)\right]\,,\\
%@@@@@@@@@@@@@@@@@@@@@@@@@@ C BEGINS @@@@@@@@@@@@@@@@@@@@@@@@@@@@@@@@@
\mathcal{C} &= -\frac{2}{7}+ \frac{6}{7}\,\nu \nonumber\\
&+\frac{1}{c^2}\left[v^2\, \left( -\frac{13}{63} + \frac{101 }{63}\,\nu
- \frac{209\,}{63}\,\nu^2 \right) \right.\nonumber\\
&\qquad \left.+\frac{G m}{r}\,\left( -\frac{155}{108} + \frac{4057}{756}\,\nu
+ \frac{209}{108}\,\nu^2 \right)\right] \nonumber\\
&+\frac{1}{c^4}\left[v^2\,\frac{G m}{r}\, \left( - \frac{2839}{1386}+
\frac{237893}{16632}\,\nu - \frac{188063}{8316}\,\nu^2
-\frac{58565}{4158}\,\nu^3 \right)\right.\nonumber\\
&\qquad+\,\left(\frac{G m}{r}\right)^2\,\left( -\frac{12587}{41580}+
\frac{406333 }{16632}\,\nu - \frac{2713}{396}\,\nu^2
+\frac{4441}{2772}\,\nu^3 \right)\nonumber\\
&\qquad+v^4\,\left( -\frac{457}{2772}+ \frac{6103 }{2772}\,\nu -
\frac{13693}{1386}\,\nu^2+\frac{40687}{2772}\,\nu^3 \right)\nonumber\\
&\qquad+\left. {\dot{r}^2\,\frac{G m}{r}\,\left( \frac{305}{5544} +
\frac{3233}{5544}\,\nu - \frac{8611}{5544}\,\nu^2 -
\frac{895}{154}\,\nu^3\right) }\right]\,.
\end{align}\end{subequations}
Those expressions were already given at 3PN order in Eqs.~(5.10) of
\cite{BI04mult}. The constants $r_0$ and $r'_0$ match the ones introduced in
Eqs.~\eqref{eq:AB}. The parts $\mathcal{A}_\text{RR}$,
$\mathcal{B}_\text{RR}$, and $\mathcal{C}_\text{RR}$ corresponding to the
radiation-reaction contributions and containing the new 3.5PN terms are:
\begin{subequations}\label{eq:ABCRR}\begin{align}
%@@@@@@@@@@@@@@@@@@@@@@@@@@ ARR BEGINS @@@@@@@@@@@@@@@@@@@@@@@@@@@@@@@@@
\mathcal{A}_\text{RR} =& 1 + \frac{1}{c^2}\left[
  \frac{Gm}{r}\left(-\frac{415}{54} - \frac{28}{405}\nu\right) +
  v^2\left(\frac{373}{54} - \frac{13543}{540}\nu\right) + \dot{r}^2\left(-5 +
    \frac{2171}{108}\nu\right)\right]\,, 
\\ 
%@@@@@@@@@@@@@@@@@@@@@@@@@@ BRR BEGINS @@@@@@@@@@@@@@@@@@@@@@@@@@@@@@@@@
\mathcal{B}_\text{RR} =& \frac{1}{c^2}\left[
  \frac{Gm}{r}\left(-\frac{664}{63}+\frac{104}{9}\nu\right) +
  v^2\left(-\frac{8}{21}+\frac{32}{21}\nu\right)\right]\,,
\\ 
%@@@@@@@@@@@@@@@@@@@@@@@@@@ CRR BEGINS @@@@@@@@@@@@@@@@@@@@@@@@@@@@@@@@@
\mathcal{C}_\text{RR} =& 1 + \frac{1}{c^2}\left[
  \frac{Gm}{r}\left(- \frac{629}{135} + \frac{871}{405}\nu\right) +
  v^2\left(\frac{226}{135} -\frac{119}{45}\nu\right)\right.\nonumber\\
& \qquad \left.+\dot{r}^2\left(\frac{1}{12}-\frac{13}{6}\nu\right)+\frac{r
    v^4}{G m}\left(\frac{1}{45}-\frac{4}{45}\nu\right)\right]\,.
\end{align}\end{subequations}
Note that we defined $\mathcal{A}_\text{RR}$, $\mathcal{B}_\text{RR}$, and
$\mathcal{C}_\text{RR}$ as being dimensionless, but that
$\mathcal{A}_\text{RR}$ and $\mathcal{C}_\text{RR}$ start at 2.5PN order,
while $\mathcal{B}_\text{RR}$ only starts at 3.5PN order. The expressions
\eqref{eq:ABCcons}--\eqref{eq:ABCRR} reduce to
Eqs.~\eqref{eq:AB}--\eqref{eq:C} in the case of quasi-circular orbits. Notice
also that we replaced the Hadamard regularization ambiguities $\xi$, $\zeta$
and $\kappa$ in the 3PN terms by their values, which can be found for instance
in Ref.~\cite{BDEI05dr}.

\bibliography{ListeRef}

%Merlin.mbs v4.21 2009-07-09.
\begin{thebibliography}{10}%
\makeatletter
\providecommand \@ifxundefined [1]{%
 \ifx #1\undefined \expandafter \@firstoftwo
 \else \expandafter \@secondoftwo
\fi
}%
\providecommand \@ifnum [1]{%
 \ifnum #1\expandafter \@firstoftwo
 \else \expandafter \@secondoftwo
\fi
}%
\providecommand \enquote [1]{``#1''}%
\providecommand \bibnamefont  [1]{#1}%
\providecommand \bibfnamefont [1]{#1}%
\providecommand \citenamefont [1]{#1}%
\providecommand\href[0]{\@sanitize\@href}%
\providecommand\@href[1]{\endgroup\@@startlink{#1}\endgroup\@@href}%
\providecommand\@@href[1]{#1\@@endlink}%
\providecommand \@sanitize [0]{\begingroup\catcode`\&12\catcode`\#12\relax}%
\@ifxundefined \pdfoutput {\@firstoftwo}{%
 \@ifnum{\z@=\pdfoutput}{\@firstoftwo}{\@secondoftwo}%
}{%
 \providecommand\@@startlink[1]{\leavevmode\special{html:<a href="#1">}}%
 \providecommand\@@endlink[0]{\special{html:</a>}}%
}{%
 \providecommand\@@startlink[1]{%
  \leavevmode
  \pdfstartlink
   attr{/Border[0 0 1 ]/H/I/C[0 1 1]}%
   user{/Subtype/Link/A<</Type/Action/S/URI/URI(#1)>>}%
  \relax
 }%
 \providecommand\@@endlink[0]{\pdfendlink}%
}%
\providecommand \url  [0]{\begingroup\@sanitize \@url }%
\providecommand \@url [1]{\endgroup\@href {#1}{\urlprefix}}%
\providecommand \urlprefix [0]{URL }%
\providecommand \Eprint[0]{\href }%
\@ifxundefined \urlstyle {%
  \providecommand \doi [1]{doi:\discretionary{}{}{}#1}%
}{%
  \providecommand \doi [0]{doi:\discretionary{}{}{}\begingroup
  \urlstyle{rm}\Url }%
}%
\providecommand \doibase [0]{http://dx.doi.org/}%
\providecommand \Doi[1]{\href{\doibase#1}}%
\providecommand \bibAnnote [3]{%
  \BibitemShut{#1}%
  \begin{quotation}\noindent
    \textsc{Key:}\ #2\\\textsc{Annotation:}\ #3%
  \end{quotation}%
}%
\providecommand \bibAnnoteFile [2]{%
  \IfFileExists{#2}{\bibAnnote {#1} {#2} {\input{#2}}}{}%
}%
\providecommand \typeout [0]{\immediate \write \m@ne }%
\providecommand \selectlanguage [0]{\@gobble}%
\providecommand \bibinfo [0]{\@secondoftwo}%
\providecommand \bibfield [0]{\@secondoftwo}%
\providecommand \translation [1]{[#1]}%
\providecommand \BibitemOpen[0]{}%
\providecommand \bibitemStop [0]{}%
\providecommand \bibitemNoStop [0]{.\EOS\space}%
\providecommand \EOS [0]{\spacefactor3000\relax}%
\providecommand \BibitemShut [1]{\csname bibitem#1\endcsname}%
%</preamble>
\bibitem{sathyaschutz09}%
  \BibitemOpen
  \bibfield{author}{%
  \bibinfo {author} {\bibfnamefont{B.}~\bibnamefont{Sathyaprakash}}\ and\
  \bibinfo {author} {\bibfnamefont{B.~F.}\ \bibnamefont{Schutz}},\ }%
  \bibfield{journal}{%
  \bibinfo {journal} {Living Rev. Rel.}\ }%
  \textbf{\bibinfo {volume} {12}} (\bibinfo {year} {2009}),\
  \url{http://www.livingreviews.org/lrr-2009-2}%
  \bibAnnoteFile{NoStop}{sathyaschutz09}%
\bibitem{Bliving}%
  \BibitemOpen
  \bibfield{author}{%
  \bibinfo {author} {\bibfnamefont{L.}~\bibnamefont{Blanchet}},\ }%
  \bibfield{journal}{%
  \bibinfo {journal} {Living Rev. Rel.}\ }%
  \textbf{\bibinfo {volume} {9}} (\bibinfo {year} {2006}),\
  \Eprint{http://arxiv.org/abs/gr-qc/0202016}{gr-qc/0202016},\
  \url{http://www.livingreviews.org/lrr-2006-4}%
  \bibAnnoteFile{NoStop}{Bliving}%
%%CITATION = GR-QC 0202016;%%
\bibitem{Pretorius}%
  \BibitemOpen
  \bibfield{author}{%
  \bibinfo {author} {\bibfnamefont{F.}~\bibnamefont{Pretorius}},\ }%
  \bibfield{journal}{%
  \bibinfo {journal} {Phys. Rev. Lett.}\ }%
  \textbf{\bibinfo {volume} {95}},\ \bibinfo {pages} {121101} (\bibinfo {year}
  {2005}),\ \Eprint{http://arxiv.org/abs/gr-qc/0507014}{gr-qc/0507014}%
  \bibAnnoteFile{NoStop}{Pretorius}%
\bibitem{Baker}%
  \BibitemOpen
  \bibfield{author}{%
  \bibinfo {author} {\bibfnamefont{J.~G.}\ \bibnamefont{Baker}}, \bibinfo
  {author} {\bibfnamefont{J.}~\bibnamefont{Centrella}}, \bibinfo {author}
  {\bibfnamefont{D.-I.}\ \bibnamefont{Choi}}, \bibinfo {author}
  {\bibfnamefont{M.}~\bibnamefont{Koppitz}},\ and\ \bibinfo {author}
  {\bibfnamefont{J.}~\bibnamefont{van Meter}},\ }%
  \bibfield{journal}{%
  \bibinfo {journal} {Phys. Rev. Lett.}\ }%
  \textbf{\bibinfo {volume} {96}},\ \bibinfo {pages} {111102} (\bibinfo {year}
  {2006}),\ \Eprint{http://arxiv.org/abs/gr-qc/0511103}{gr-qc/0511103}%
  \bibAnnoteFile{NoStop}{Baker}%
\bibitem{Campanelli}%
  \BibitemOpen
  \bibfield{author}{%
  \bibinfo {author} {\bibfnamefont{M.}~\bibnamefont{Campanelli}}, \bibinfo
  {author} {\bibfnamefont{C.~O.}\ \bibnamefont{Lousto}}, \bibinfo {author}
  {\bibfnamefont{P.}~\bibnamefont{Marronetti}},\ and\ \bibinfo {author}
  {\bibfnamefont{Y.}~\bibnamefont{Zwlochower}},\ }%
  \bibfield{journal}{%
  \bibinfo {journal} {Phys. Rev. Lett.}\ }%
  \textbf{\bibinfo {volume} {96}},\ \bibinfo {pages} {111101} (\bibinfo {year}
  {2006}),\ \Eprint{http://arxiv.org/abs/gr-qc/0511048}{gr-qc/0511048}%
  \bibAnnoteFile{NoStop}{Campanelli}%
\bibitem{BCPZ}%
  \BibitemOpen
  \bibfield{author}{%
  \bibinfo {author} {\bibfnamefont{J.}~\bibnamefont{Baker}}, \bibinfo {author}
  {\bibfnamefont{M.}~\bibnamefont{Campanelli}}, \bibinfo {author}
  {\bibfnamefont{F.}~\bibnamefont{Pretorius}},\ and\ \bibinfo {author}
  {\bibfnamefont{Y.}~\bibnamefont{Zlochower}},\ }%
  \bibfield{journal}{%
  \bibinfo {journal} {Class. Quant. Grav.}\ }%
  \textbf{\bibinfo {volume} {24}},\ \bibinfo {pages} {S25} (\bibinfo {year}
  {2007})%
  \bibAnnoteFile{NoStop}{BCPZ}%
\bibitem{Hannam09}%
  \BibitemOpen
  \bibfield{author}{%
  \bibinfo {author} {\bibfnamefont{M.}~\bibnamefont{Hannam}}, \bibinfo {author}
  {\bibfnamefont{S.}~\bibnamefont{Husa}}, \bibinfo {author}
  {\bibfnamefont{J.~G.}\ \bibnamefont{Baker}}, \bibinfo {author}
  {\bibfnamefont{M.}~\bibnamefont{Boyle}}, \bibinfo {author}
  {\bibfnamefont{B.}~\bibnamefont{Bruegmann}}, \emph{et~al.},\ }%
  \bibfield{journal}{%
  \Doi{10.1103/PhysRevD.79.084025}{\bibinfo {journal} {Phys.Rev.}}\ }%
  \textbf{\bibinfo {volume} {D79}},\ \bibinfo {pages} {084025} (\bibinfo {year}
  {2009}),\ \Eprint{http://arxiv.org/abs/0901.2437}{arXiv:0901.2437 [gr-qc]}%
  \bibAnnoteFile{NoStop}{Hannam09}%
%%CITATION = ARXIV:0901.2437;%%
\bibitem{BCP07}%
  \BibitemOpen
  \bibfield{author}{%
  \bibinfo {author} {\bibfnamefont{A.}~\bibnamefont{Buonanno}}, \bibinfo
  {author} {\bibfnamefont{G.~B.}\ \bibnamefont{Cook}},\ and\ \bibinfo {author}
  {\bibfnamefont{F.}~\bibnamefont{Pretorius}},\ }%
  \bibfield{journal}{%
  \bibinfo {journal} {Phys. Rev. D}\ }%
  \textbf{\bibinfo {volume} {75}},\ \bibinfo {pages} {124018} (\bibinfo {year}
  {2007}),\ \Eprint{http://arxiv.org/abs/gr-qc/0610122}{gr-qc/0610122}%
  \bibAnnoteFile{NoStop}{BCP07}%
\bibitem{Berti}%
  \BibitemOpen
  \bibfield{author}{%
  \bibinfo {author} {\bibfnamefont{E.}~\bibnamefont{Berti}}, \bibinfo {author}
  {\bibfnamefont{V.}~\bibnamefont{Cardoso}}, \bibinfo {author}
  {\bibfnamefont{J.~A.}\ \bibnamefont{Gonzalez}}, \bibinfo {author}
  {\bibfnamefont{U.}~\bibnamefont{Sperhake}}, \bibinfo {author}
  {\bibfnamefont{M.}~\bibnamefont{Hannam}}, \bibinfo {author}
  {\bibfnamefont{S.}~\bibnamefont{Husa}},\ and\ \bibinfo {author}
  {\bibfnamefont{B.}~\bibnamefont{Bruegmann}},\ }%
  \bibfield{journal}{%
  \bibinfo {journal} {Phys. Rev. D}\ }%
  \textbf{\bibinfo {volume} {76}},\ \bibinfo {pages} {064034} (\bibinfo {year}
  {2007})%
  \bibAnnoteFile{NoStop}{Berti}%
\bibitem{Jena}%
  \BibitemOpen
  \bibfield{author}{%
  \bibinfo {author} {\bibfnamefont{M.}~\bibnamefont{Hannam}}, \bibinfo {author}
  {\bibfnamefont{S.}~\bibnamefont{Husa}}, \bibinfo {author}
  {\bibfnamefont{J.~A.}\ \bibnamefont{Gonz\'alez}}, \bibinfo {author}
  {\bibfnamefont{U.}~\bibnamefont{Sperhake}},\ and\ \bibinfo {author}
  {\bibfnamefont{B.}~\bibnamefont{Br\"ugmann}},\ }%
  \bibfield{journal}{%
  \Doi{10.1103/PhysRevD.77.044020}{\bibinfo {journal} {Phys. Rev. D}}\ }%
  \textbf{\bibinfo {volume} {77}},\ \bibinfo {pages} {044020} (\bibinfo {month}
  {Feb}\ \bibinfo {year} {2008}),\
  \url{http://link.aps.org/doi/10.1103/PhysRevD.77.044020}%
  \bibAnnoteFile{NoStop}{Jena}%
\bibitem{Boyle}%
  \BibitemOpen
  \bibfield{author}{%
  \bibinfo {author} {\bibfnamefont{M.}~\bibnamefont{Boyle}}, \bibinfo {author}
  {\bibfnamefont{D.~A.}\ \bibnamefont{Brown}}, \bibinfo {author}
  {\bibfnamefont{L.~E.}\ \bibnamefont{Kidder}}, \bibinfo {author}
  {\bibfnamefont{A.~H.}\ \bibnamefont{Mroue}}, \bibinfo {author}
  {\bibfnamefont{H.~P.}\ \bibnamefont{Pfeiffer}}, \bibinfo {author}
  {\bibfnamefont{M.~A.}\ \bibnamefont{Scheel}}, \bibinfo {author}
  {\bibfnamefont{G.~B.}\ \bibnamefont{Cook}},\ and\ \bibinfo {author}
  {\bibfnamefont{S.~A.}\ \bibnamefont{Teukolsky}},\ }%
  \bibfield{journal}{%
  \bibinfo {journal} {Phys. Rev. D}\ }%
  \textbf{\bibinfo {volume} {76}},\ \bibinfo {pages} {124038} (\bibinfo {year}
  {2007}),\ \Eprint{http://arxiv.org/abs/arXiv:0710.0158
  [gr-qc]}{arXiv:0710.0158 [gr-qc]}%
  \bibAnnoteFile{NoStop}{Boyle}%
\bibitem{Ajith08}%
  \BibitemOpen
  \bibfield{author}{%
  \bibinfo {author} {\bibfnamefont{P.}~\bibnamefont{Ajith}} \emph{et~al.},\ }%
  \bibfield{journal}{%
  \bibinfo {journal} {Phys. Rev. D}\ }%
  \textbf{\bibinfo {volume} {77}},\ \bibinfo {pages} {104017} (\bibinfo {year}
  {2008}),\ \bibinfo {note} {{E}rratum: Phys. Rev. D \textbf{79}, 129901(E)
  (2009)},\ \Eprint{http://arxiv.org/abs/arXiv:0710.2335
  [gr-qc]}{arXiv:0710.2335 [gr-qc]}%
  \bibAnnoteFile{NoStop}{Ajith08}%
\bibitem{Buo09}%
  \BibitemOpen
  \bibfield{author}{%
  \bibinfo {author} {\bibfnamefont{A.}~\bibnamefont{Buonanno}}, \bibinfo
  {author} {\bibfnamefont{Y.}~\bibnamefont{Pan}}, \bibinfo {author}
  {\bibfnamefont{H.~P.}\ \bibnamefont{Pfeiffer}}, \bibinfo {author}
  {\bibfnamefont{M.~A.}\ \bibnamefont{Scheel}}, \bibinfo {author}
  {\bibfnamefont{L.~T.}\ \bibnamefont{Buchman}}, \emph{et~al.},\ }%
  \bibfield{journal}{%
  \Doi{10.1103/PhysRevD.79.124028}{\bibinfo {journal} {Phys.Rev.}}\ }%
  \textbf{\bibinfo {volume} {D79}},\ \bibinfo {pages} {124028} (\bibinfo {year}
  {2009}),\ \Eprint{http://arxiv.org/abs/0902.0790}{arXiv:0902.0790 [gr-qc]}%
  \bibAnnoteFile{NoStop}{Buo09}%
\bibitem{Pan09}%
  \BibitemOpen
  \bibfield{author}{%
  \bibinfo {author} {\bibfnamefont{Y.}~\bibnamefont{Pan}}, \bibinfo {author}
  {\bibfnamefont{A.}~\bibnamefont{Buonanno}}, \bibinfo {author}
  {\bibfnamefont{L.~T.}\ \bibnamefont{Buchman}}, \bibinfo {author}
  {\bibfnamefont{T.}~\bibnamefont{Chu}}, \bibinfo {author}
  {\bibfnamefont{L.~E.}\ \bibnamefont{Kidder}}, \emph{et~al.},\ }%
  \bibfield{journal}{%
  \Doi{10.1103/PhysRevD.81.084041}{\bibinfo {journal} {Phys.Rev.}}\ }%
  \textbf{\bibinfo {volume} {D81}},\ \bibinfo {pages} {084041} (\bibinfo {year}
  {2010}),\ \Eprint{http://arxiv.org/abs/0912.3466}{arXiv:0912.3466 [gr-qc]}%
  \bibAnnoteFile{NoStop}{Pan09}%
\bibitem{PanBFRT11}%
  \BibitemOpen
  \bibfield{author}{%
  \bibinfo {author} {\bibfnamefont{Y.}~\bibnamefont{Pan}}, \bibinfo {author}
  {\bibfnamefont{A.}~\bibnamefont{Buonanno}}, \bibinfo {author}
  {\bibfnamefont{R.}~\bibnamefont{Fujita}}, \bibinfo {author}
  {\bibfnamefont{E.}~\bibnamefont{Racine}},\ and\ \bibinfo {author}
  {\bibfnamefont{H.}~\bibnamefont{Tagoshi}},\ }%
  \bibfield{journal}{%
  \Doi{10.1103/PhysRevD.83.064003}{\bibinfo {journal} {Phys. Rev. D}}\ }%
  \textbf{\bibinfo {volume} {83}},\ \bibinfo {pages} {064003} (\bibinfo {month}
  {Mar}\ \bibinfo {year} {2011}),\
  \Eprint{http://arxiv.org/abs/1006.0431}{arXiv:1006.0431},\
  \url{http://link.aps.org/doi/10.1103/PhysRevD.83.064003}%
  \bibAnnoteFile{NoStop}{PanBFRT11}%
\bibitem{BIWW96}%
  \BibitemOpen
  \bibfield{author}{%
  \bibinfo {author} {\bibfnamefont{L.}~\bibnamefont{Blanchet}}, \bibinfo
  {author} {\bibfnamefont{B.~R.}\ \bibnamefont{Iyer}}, \bibinfo {author}
  {\bibfnamefont{C.~M.}\ \bibnamefont{Will}},\ and\ \bibinfo {author}
  {\bibfnamefont{A.~G.}\ \bibnamefont{Wiseman}},\ }%
  \bibfield{journal}{%
  \bibinfo {journal} {Class. Quant. Grav.}\ }%
  \textbf{\bibinfo {volume} {13}},\ \bibinfo {pages} {575} (\bibinfo {year}
  {1996}),\ \Eprint{http://arxiv.org/abs/gr-qc/9602024}{gr-qc/9602024}%
  \bibAnnoteFile{NoStop}{BIWW96}%
%%CITATION = GR-QC 9602024;%%
\bibitem{ABIQ04}%
  \BibitemOpen
  \bibfield{author}{%
  \bibinfo {author} {\bibfnamefont{K.}~\bibnamefont{Arun}}, \bibinfo {author}
  {\bibfnamefont{L.}~\bibnamefont{Blanchet}}, \bibinfo {author}
  {\bibfnamefont{B.~R.}\ \bibnamefont{Iyer}},\ and\ \bibinfo {author}
  {\bibfnamefont{M.~S.~S.}\ \bibnamefont{Qusailah}},\ }%
  \bibfield{journal}{%
  \bibinfo {journal} {Class. Quant. Grav.}\ }%
  \textbf{\bibinfo {volume} {21}},\ \bibinfo {pages} {3771} (\bibinfo {year}
  {2004}),\ \bibinfo {note} {erratum \textit{Class. Quant. Grav.}, 22:3115,
  2005},\ \Eprint{http://arxiv.org/abs/gr-qc/0404085}{gr-qc/0404085}%
  \bibAnnoteFile{NoStop}{ABIQ04}%
\bibitem{KBI07}%
  \BibitemOpen
  \bibfield{author}{%
  \bibinfo {author} {\bibfnamefont{L.}~\bibnamefont{Kidder}}, \bibinfo {author}
  {\bibfnamefont{L.}~\bibnamefont{Blanchet}},\ and\ \bibinfo {author}
  {\bibfnamefont{B.~R.}\ \bibnamefont{Iyer}},\ }%
  \bibfield{journal}{%
  \bibinfo {journal} {Class. Quant. Grav.}\ }%
  \textbf{\bibinfo {volume} {24}},\ \bibinfo {pages} {5307} (\bibinfo {year}
  {2007}),\ \Eprint{http://arxiv.org/abs/arXiv:0706.0726}{arXiv:0706.0726}%
  \bibAnnoteFile{NoStop}{KBI07}%
\bibitem{K08}%
  \BibitemOpen
  \bibfield{author}{%
  \bibinfo {author} {\bibfnamefont{L.~E.}\ \bibnamefont{Kidder}},\ }%
  \bibfield{journal}{%
  \Doi{10.1103/PhysRevD.77.044016}{\bibinfo {journal} {Phys. Rev. D}}\ }%
  \textbf{\bibinfo {volume} {77}},\ \bibinfo {pages} {044016} (\bibinfo {year}
  {2008}),\ \Eprint{http://arxiv.org/abs/0710.0614}{arXiv:0710.0614 [gr-qc]}%
  \bibAnnoteFile{NoStop}{K08}%
\bibitem{BFIS08}%
  \BibitemOpen
  \bibfield{author}{%
  \bibinfo {author} {\bibfnamefont{L.}~\bibnamefont{Blanchet}}, \bibinfo
  {author} {\bibfnamefont{G.}~\bibnamefont{Faye}}, \bibinfo {author}
  {\bibfnamefont{B.~R.}\ \bibnamefont{Iyer}},\ and\ \bibinfo {author}
  {\bibfnamefont{S.}~\bibnamefont{Sinha}},\ }%
  \bibfield{journal}{%
  \bibinfo {journal} {Class. Quant. Grav.}\ }%
  \textbf{\bibinfo {volume} {25}},\ \bibinfo {pages} {165003} (\bibinfo {year}
  {2008})%
  \bibAnnoteFile{NoStop}{BFIS08}%
\bibitem{AISSV}%
  \BibitemOpen
  \bibfield{author}{%
  \bibinfo {author} {\bibfnamefont{K.~G.}\ \bibnamefont{Arun}}, \bibinfo
  {author} {\bibfnamefont{B.~R.}\ \bibnamefont{Iyer}}, \bibinfo {author}
  {\bibfnamefont{B.~S.}\ \bibnamefont{Sathyaprakash}}, \bibinfo {author}
  {\bibfnamefont{S.}~\bibnamefont{Sinha}},\ and\ \bibinfo {author}
  {\bibfnamefont{C.}~\bibnamefont{Van Den~Broeck}},\ }%
  \bibfield{journal}{%
  \Doi{10.1103/PhysRevD.76.104016}{\bibinfo {journal} {Phys. Rev. D}}\ }%
  \textbf{\bibinfo {volume} {76}},\ \bibinfo {pages} {104016} (\bibinfo {month}
  {Nov}\ \bibinfo {year} {2007}),\
  \url{http://link.aps.org/doi/10.1103/PhysRevD.76.104016}%
  \bibAnnoteFile{NoStop}{AISSV}%
\bibitem{TS08}%
  \BibitemOpen
  \bibfield{author}{%
  \bibinfo {author} {\bibfnamefont{M.}~\bibnamefont{Trias}}\ and\ \bibinfo
  {author} {\bibfnamefont{A.~M.}\ \bibnamefont{Sintes}},\ }%
  \bibfield{journal}{%
  \Doi{10.1103/PhysRevD.77.024030}{\bibinfo {journal} {Phys. Rev. D}}\ }%
  \textbf{\bibinfo {volume} {77}},\ \bibinfo {pages} {024030} (\bibinfo {month}
  {Jan}\ \bibinfo {year} {2008}),\
  \Eprint{http://arxiv.org/abs/arXiv:0707.4434}{arXiv:0707.4434},\
  \url{http://link.aps.org/doi/10.1103/PhysRevD.77.024030}%
  \bibAnnoteFile{NoStop}{TS08}%
\bibitem{TSasa94}%
  \BibitemOpen
  \bibfield{author}{%
  \bibinfo {author} {\bibfnamefont{H.}~\bibnamefont{Tagoshi}}\ and\ \bibinfo
  {author} {\bibfnamefont{M.}~\bibnamefont{Sasaki}},\ }%
  \bibfield{journal}{%
  \bibinfo {journal} {Prog. Theor. Phys.}\ }%
  \textbf{\bibinfo {volume} {92}},\ \bibinfo {pages} {745} (\bibinfo {year}
  {1994})%
  \bibAnnoteFile{NoStop}{TSasa94}%
\bibitem{FI10}%
  \BibitemOpen
  \bibfield{author}{%
  \bibinfo {author} {\bibfnamefont{R.}~\bibnamefont{Fujita}}\ and\ \bibinfo
  {author} {\bibfnamefont{B.~R.}\ \bibnamefont{Iyer}},\ }%
  \bibfield{journal}{%
  \Doi{10.1103/PhysRevD.82.044051}{\bibinfo {journal} {Phys. Rev. D}}\ }%
  \textbf{\bibinfo {volume} {82}},\ \bibinfo {pages} {044051} (\bibinfo {month}
  {Aug}\ \bibinfo {year} {2010}),\
  \Eprint{http://arxiv.org/abs/1005.2266}{arXiv:1005.2266},\
  \url{http://link.aps.org/doi/10.1103/PhysRevD.82.044051}%
  \bibAnnoteFile{NoStop}{FI10}%
\bibitem{BD86}%
  \BibitemOpen
  \bibfield{author}{%
  \bibinfo {author} {\bibfnamefont{L.}~\bibnamefont{Blanchet}}\ and\ \bibinfo
  {author} {\bibfnamefont{T.}~\bibnamefont{Damour}},\ }%
  \bibfield{journal}{%
  \bibinfo {journal} {Phil. Trans. Roy. Soc. Lond. A}\ }%
  \textbf{\bibinfo {volume} {320}},\ \bibinfo {pages} {379} (\bibinfo {year}
  {1986})%
  \bibAnnoteFile{NoStop}{BD86}%
%%CITATION = PTRSA,A320,379;%%
\bibitem{BD92}%
  \BibitemOpen
  \bibfield{author}{%
  \bibinfo {author} {\bibfnamefont{L.}~\bibnamefont{Blanchet}}\ and\ \bibinfo
  {author} {\bibfnamefont{T.}~\bibnamefont{Damour}},\ }%
  \bibfield{journal}{%
  \bibinfo {journal} {Phys. Rev. D}\ }%
  \textbf{\bibinfo {volume} {46}},\ \bibinfo {pages} {4304} (\bibinfo {year}
  {1992})%
  \bibAnnoteFile{NoStop}{BD92}%
%%CITATION = PHRVA,D46,4304;%%
\bibitem{B98quad}%
  \BibitemOpen
  \bibfield{author}{%
  \bibinfo {author} {\bibfnamefont{L.}~\bibnamefont{Blanchet}},\ }%
  \bibfield{journal}{%
  \bibinfo {journal} {Class. Quant. Grav.}\ }%
  \textbf{\bibinfo {volume} {15}},\ \bibinfo {pages} {89} (\bibinfo {year}
  {1998}),\ \Eprint{http://arxiv.org/abs/gr-qc/9710037}{gr-qc/9710037}%
  \bibAnnoteFile{NoStop}{B98quad}%
%%CITATION = GR-QC 9710037;%%
\bibitem{B98tail}%
  \BibitemOpen
  \bibfield{author}{%
  \bibinfo {author} {\bibfnamefont{L.}~\bibnamefont{Blanchet}},\ }%
  \bibfield{journal}{%
  \bibinfo {journal} {Class. Quant. Grav.}\ }%
  \textbf{\bibinfo {volume} {15}},\ \bibinfo {pages} {113} (\bibinfo {year}
  {1998}),\ \bibinfo {note} {erratum \textit{Class. Quant. Grav.}, 22:3381,
  2005},\ \Eprint{http://arxiv.org/abs/gr-qc/9710038}{gr-qc/9710038}%
  \bibAnnoteFile{NoStop}{B98tail}%
%%CITATION = GR-QC 9710038;%%
\bibitem{B98mult}%
  \BibitemOpen
  \bibfield{author}{%
  \bibinfo {author} {\bibfnamefont{L.}~\bibnamefont{Blanchet}},\ }%
  \bibfield{journal}{%
  \bibinfo {journal} {Class. Quant. Grav.}\ }%
  \textbf{\bibinfo {volume} {15}},\ \bibinfo {pages} {1971} (\bibinfo {year}
  {1998}),\ \Eprint{http://arxiv.org/abs/gr-qc/9801101}{gr-qc/9801101}%
  \bibAnnoteFile{NoStop}{B98mult}%
%%CITATION = GR-QC 9801101;%%
\bibitem{BIJ02}%
  \BibitemOpen
  \bibfield{author}{%
  \bibinfo {author} {\bibfnamefont{L.}~\bibnamefont{Blanchet}}, \bibinfo
  {author} {\bibfnamefont{B.~R.}\ \bibnamefont{Iyer}},\ and\ \bibinfo {author}
  {\bibfnamefont{B.}~\bibnamefont{Joguet}},\ }%
  \bibfield{journal}{%
  \bibinfo {journal} {Phys. Rev. D}\ }%
  \textbf{\bibinfo {volume} {65}},\ \bibinfo {pages} {064005} (\bibinfo {year}
  {2002}),\ \bibinfo {note} {erratum \textit{Phys. Rev. D}, 71:129903(E),
  2005},\ \Eprint{http://arxiv.org/abs/gr-qc/0105098}{gr-qc/0105098}%
  \bibAnnoteFile{NoStop}{BIJ02}%
\bibitem{BI04mult}%
  \BibitemOpen
  \bibfield{author}{%
  \bibinfo {author} {\bibfnamefont{L.}~\bibnamefont{Blanchet}}\ and\ \bibinfo
  {author} {\bibfnamefont{B.~R.}\ \bibnamefont{Iyer}},\ }%
  \bibfield{journal}{%
  \bibinfo {journal} {Phys. Rev. D}\ }%
  \textbf{\bibinfo {volume} {71}},\ \bibinfo {pages} {024004} (\bibinfo {year}
  {2004}),\ \Eprint{http://arxiv.org/abs/gr-qc/0409094}{gr-qc/0409094}%
  \bibAnnoteFile{NoStop}{BI04mult}%
\bibitem{BFreg}%
  \BibitemOpen
  \bibfield{author}{%
  \bibinfo {author} {\bibfnamefont{L.}~\bibnamefont{Blanchet}}\ and\ \bibinfo
  {author} {\bibfnamefont{G.}~\bibnamefont{Faye}},\ }%
  \bibfield{journal}{%
  \bibinfo {journal} {J. Math. Phys.}\ }%
  \textbf{\bibinfo {volume} {41}},\ \bibinfo {pages} {7675} (\bibinfo {year}
  {2000}),\ \Eprint{http://arxiv.org/abs/gr-qc/0004008}{gr-qc/0004008}%
  \bibAnnoteFile{NoStop}{BFreg}%
%%CITATION = GR-QC 0004008;%%
\bibitem{BF00}%
  \BibitemOpen
  \bibfield{author}{%
  \bibinfo {author} {\bibfnamefont{L.}~\bibnamefont{Blanchet}}\ and\ \bibinfo
  {author} {\bibfnamefont{G.}~\bibnamefont{Faye}},\ }%
  \bibfield{journal}{%
  \bibinfo {journal} {Phys. Lett. A}\ }%
  \textbf{\bibinfo {volume} {271}},\ \bibinfo {pages} {58} (\bibinfo {year}
  {2000}),\ \Eprint{http://arxiv.org/abs/gr-qc/0004009}{gr-qc/0004009}%
  \bibAnnoteFile{NoStop}{BF00}%
%%CITATION = GR-QC 0004009;%%
\bibitem{BFeom}%
  \BibitemOpen
  \bibfield{author}{%
  \bibinfo {author} {\bibfnamefont{L.}~\bibnamefont{Blanchet}}\ and\ \bibinfo
  {author} {\bibfnamefont{G.}~\bibnamefont{Faye}},\ }%
  \bibfield{journal}{%
  \bibinfo {journal} {Phys. Rev. D}\ }%
  \textbf{\bibinfo {volume} {63}},\ \bibinfo {pages} {062005} (\bibinfo {year}
  {2001}),\ \Eprint{http://arxiv.org/abs/gr-qc/0007051}{gr-qc/0007051}%
  \bibAnnoteFile{NoStop}{BFeom}%
%%CITATION = GR-QC 0007051;%%
\bibitem{BDE04}%
  \BibitemOpen
  \bibfield{author}{%
  \bibinfo {author} {\bibfnamefont{L.}~\bibnamefont{Blanchet}}, \bibinfo
  {author} {\bibfnamefont{T.}~\bibnamefont{Damour}},\ and\ \bibinfo {author}
  {\bibfnamefont{G.}~\bibnamefont{Esposito-Far{\`e}se}},\ }%
  \bibfield{journal}{%
  \bibinfo {journal} {Phys. Rev. D}\ }%
  \textbf{\bibinfo {volume} {69}},\ \bibinfo {pages} {124007} (\bibinfo {year}
  {2004}),\ \Eprint{http://arxiv.org/abs/gr-qc/0311052}{gr-qc/0311052}%
  \bibAnnoteFile{NoStop}{BDE04}%
%%CITATION = GR-QC 0311052;%%
\bibitem{IFA01}%
  \BibitemOpen
  \bibfield{author}{%
  \bibinfo {author} {\bibfnamefont{Y.}~\bibnamefont{Itoh}}, \bibinfo {author}
  {\bibfnamefont{T.}~\bibnamefont{Futamase}},\ and\ \bibinfo {author}
  {\bibfnamefont{H.}~\bibnamefont{Asada}},\ }%
  \bibfield{journal}{%
  \bibinfo {journal} {Phys. Rev. D}\ }%
  \textbf{\bibinfo {volume} {63}},\ \bibinfo {pages} {064038} (\bibinfo {year}
  {2001})%
  \bibAnnoteFile{NoStop}{IFA01}%
\bibitem{itoh1}%
  \BibitemOpen
  \bibfield{author}{%
  \bibinfo {author} {\bibfnamefont{Y.}~\bibnamefont{Itoh}}\ and\ \bibinfo
  {author} {\bibfnamefont{T.}~\bibnamefont{Futamase}},\ }%
  \bibfield{journal}{%
  \bibinfo {journal} {Phys. Rev. D}\ }%
  \textbf{\bibinfo {volume} {68}},\ \bibinfo {pages} {121501(R)} (\bibinfo
  {year} {2003}),\
  \Eprint{http://arxiv.org/abs/arXiv:gr-qc/0310028}{arXiv:gr-qc/0310028}%
  \bibAnnoteFile{NoStop}{itoh1}%
\bibitem{itoh2}%
  \BibitemOpen
  \bibfield{author}{%
  \bibinfo {author} {\bibfnamefont{Y.}~\bibnamefont{Itoh}},\ }%
  \bibfield{journal}{%
  \bibinfo {journal} {Phys. Rev. D}\ }%
  \textbf{\bibinfo {volume} {69}},\ \bibinfo {pages} {064018} (\bibinfo {year}
  {2004}),\
  \Eprint{http://arxiv.org/abs/arXiv:gr-qc/0310029}{arXiv:gr-qc/0310029}%
  \bibAnnoteFile{NoStop}{itoh2}%
\bibitem{FS11}%
  \BibitemOpen
  \bibfield{author}{%
  \bibinfo {author} {\bibfnamefont{S.}~\bibnamefont{Foffa}}\ and\ \bibinfo
  {author} {\bibfnamefont{R.}~\bibnamefont{Sturani}},\ }%
  \bibfield{journal}{%
  \Doi{10.1103/PhysRevD.84.044031}{\bibinfo {journal} {Phys. Rev. D}}\ }%
  \textbf{\bibinfo {volume} {84}},\ \bibinfo {pages} {044031} (\bibinfo {month}
  {Aug}\ \bibinfo {year} {2011}),\ \Eprint{http://arxiv.org/abs/1104.1122
  [gr-qc]}{arXiv:1104.1122 [gr-qc]},\
  \url{http://link.aps.org/doi/10.1103/PhysRevD.84.044031}%
  \bibAnnoteFile{NoStop}{FS11}%
\bibitem{JaraS98}%
  \BibitemOpen
  \bibfield{author}{%
  \bibinfo {author} {\bibfnamefont{P.}~\bibnamefont{Jaranowski}}\ and\ \bibinfo
  {author} {\bibfnamefont{G.}~\bibnamefont{Sch\"afer}},\ }%
  \bibfield{journal}{%
  \bibinfo {journal} {Phys. Rev. D}\ }%
  \textbf{\bibinfo {volume} {57}},\ \bibinfo {pages} {7274} (\bibinfo {year}
  {1998}),\
  \Eprint{http://arxiv.org/abs/arXiv:gr-qc/9712075}{arXiv:gr-qc/9712075}%
  \bibAnnoteFile{NoStop}{JaraS98}%
\bibitem{JaraS99}%
  \BibitemOpen
  \bibfield{author}{%
  \bibinfo {author} {\bibfnamefont{P.}~\bibnamefont{Jaranowski}}\ and\ \bibinfo
  {author} {\bibfnamefont{G.}~\bibnamefont{Sch\"afer}},\ }%
  \bibfield{journal}{%
  \bibinfo {journal} {Phys. Rev. D}\ }%
  \textbf{\bibinfo {volume} {60}},\ \bibinfo {pages} {124003} (\bibinfo {year}
  {1999})%
  \bibAnnoteFile{NoStop}{JaraS99}%
\bibitem{DJSdim}%
  \BibitemOpen
  \bibfield{author}{%
  \bibinfo {author} {\bibfnamefont{T.}~\bibnamefont{Damour}}, \bibinfo {author}
  {\bibfnamefont{P.}~\bibnamefont{Jaranowski}},\ and\ \bibinfo {author}
  {\bibfnamefont{G.}~\bibnamefont{Sch\"afer}},\ }%
  \bibfield{journal}{%
  \bibinfo {journal} {Phys. Lett. B}\ }%
  \textbf{\bibinfo {volume} {513}},\ \bibinfo {pages} {147} (\bibinfo {year}
  {2001}),\ \Eprint{http://arxiv.org/abs/gr-qc/0105038}{gr-qc/0105038}%
  \bibAnnoteFile{NoStop}{DJSdim}%
\bibitem{PW02}%
  \BibitemOpen
  \bibfield{author}{%
  \bibinfo {author} {\bibfnamefont{M.}~\bibnamefont{Pati}}\ and\ \bibinfo
  {author} {\bibfnamefont{C.}~\bibnamefont{Will}},\ }%
  \bibfield{journal}{%
  \bibinfo {journal} {Phys. Rev. D}\ }%
  \textbf{\bibinfo {volume} {65}},\ \bibinfo {pages} {104008} (\bibinfo {year}
  {2002}),\ \Eprint{http://arxiv.org/abs/gr-qc/0201001}{gr-qc/0201001}%
  \bibAnnoteFile{NoStop}{PW02}%
\bibitem{NB05}%
  \BibitemOpen
  \bibfield{author}{%
  \bibinfo {author} {\bibfnamefont{S.}~\bibnamefont{Nissanke}}\ and\ \bibinfo
  {author} {\bibfnamefont{L.}~\bibnamefont{Blanchet}},\ }%
  \bibfield{journal}{%
  \bibinfo {journal} {Class. Quant. Grav.}\ }%
  \textbf{\bibinfo {volume} {22}},\ \bibinfo {pages} {1007} (\bibinfo {year}
  {2005}),\ \Eprint{http://arxiv.org/abs/gr-qc/0412018}{gr-qc/0412018}%
  \bibAnnoteFile{NoStop}{NB05}%
\bibitem{Th80}%
  \BibitemOpen
  \bibfield{author}{%
  \bibinfo {author} {\bibfnamefont{K.}~\bibnamefont{Thorne}},\ }%
  \bibfield{journal}{%
  \bibinfo {journal} {Rev. Mod. Phys.}\ }%
  \textbf{\bibinfo {volume} {52}},\ \bibinfo {pages} {299} (\bibinfo {year}
  {1980})%
  \bibAnnoteFile{NoStop}{Th80}%
\bibitem{GMNRS67}%
  \BibitemOpen
  \bibfield{author}{%
  \bibinfo {author} {\bibfnamefont{J.}~\bibnamefont{Goldberg}}, \bibinfo
  {author} {\bibfnamefont{A.}~\bibnamefont{MacFarlane}}, \bibinfo {author}
  {\bibfnamefont{E.}~\bibnamefont{Newman}}, \bibinfo {author}
  {\bibfnamefont{F.}~\bibnamefont{Rohrlich}},\ and\ \bibinfo {author}
  {\bibfnamefont{E.}~\bibnamefont{Sudarshan}},\ }%
  \bibfield{journal}{%
  \Doi{doi:10.1063/1.1705135}{\bibinfo {journal} {J. Math. Phys.}}\ }%
  \textbf{\bibinfo {volume} {8}},\ \bibinfo {pages} {2155} (\bibinfo {year}
  {1967})%
  \bibAnnoteFile{NoStop}{GMNRS67}%
\bibitem{xtensor}%
  \BibitemOpen
  \bibfield{author}{%
  \bibinfo {author} {\bibfnamefont{J.~M.}\ \bibnamefont{Mart\'in-Garc\'ia}},
  \bibinfo {author} {\bibfnamefont{A.}~\bibnamefont{Garc\'ia-Parrado}},
  \bibinfo {author} {\bibfnamefont{A.}~\bibnamefont{Stecchina}}, \bibinfo
  {author} {\bibfnamefont{B.}~\bibnamefont{Wardell}}, \bibinfo {author}
  {\bibfnamefont{C.}~\bibnamefont{Pitrou}}, \bibinfo {author}
  {\bibfnamefont{D.}~\bibnamefont{Brizuela}}, \bibinfo {author}
  {\bibfnamefont{D.}~\bibnamefont{Yllanes}}, \bibinfo {author}
  {\bibfnamefont{G.}~\bibnamefont{Faye}}, \bibinfo {author}
  {\bibfnamefont{L.}~\bibnamefont{Stein}}, \bibinfo {author}
  {\bibfnamefont{R.}~\bibnamefont{Portugal}},\ and\ \bibinfo {author}
  {\bibfnamefont{T.}~\bibnamefont{B\"ackdahl}},\ }%
  \enquote{\bibinfo {title} {{xAct}: Efficient tensor computer algebra for
  {Mathematica}},}\  (\bibinfo {year} {GPL 2002--2012}),\ \bibinfo {note}
  {http://www.xact.es/}%
  \bibAnnoteFile{NoStop}{xtensor}%
\bibitem{Chr91}%
  \BibitemOpen
  \bibfield{author}{%
  \bibinfo {author} {\bibfnamefont{D.}~\bibnamefont{Christodoulou}},\ }%
  \bibfield{journal}{%
  \bibinfo {journal} {Phys. Rev. Lett.}\ }%
  \textbf{\bibinfo {volume} {67}},\ \bibinfo {pages} {1486} (\bibinfo {year}
  {1991})%
  \bibAnnoteFile{NoStop}{Chr91}%
\bibitem{Th92}%
  \BibitemOpen
  \bibfield{author}{%
  \bibinfo {author} {\bibfnamefont{K.}~\bibnamefont{Thorne}},\ }%
  \bibfield{journal}{%
  \bibinfo {journal} {Phys. Rev. D}\ }%
  \textbf{\bibinfo {volume} {45}},\ \bibinfo {pages} {520} (\bibinfo {year}
  {1992})%
  \bibAnnoteFile{NoStop}{Th92}%
\bibitem{WW91}%
  \BibitemOpen
  \bibfield{author}{%
  \bibinfo {author} {\bibfnamefont{A.}~\bibnamefont{Wiseman}}\ and\ \bibinfo
  {author} {\bibfnamefont{C.}~\bibnamefont{Will}},\ }%
  \bibfield{journal}{%
  \bibinfo {journal} {Phys. Rev. D}\ }%
  \textbf{\bibinfo {volume} {44}},\ \bibinfo {pages} {R2945} (\bibinfo {year}
  {1991})%
  \bibAnnoteFile{NoStop}{WW91}%
\bibitem{F09}%
  \BibitemOpen
  \bibfield{author}{%
  \bibinfo {author} {\bibfnamefont{M.}~\bibnamefont{Favata}},\ }%
  \bibfield{journal}{%
  \Doi{10.1103/PhysRevD.80.024002}{\bibinfo {journal} {Phys. Rev. D}}\ }%
  \textbf{\bibinfo {volume} {80}},\ \bibinfo {pages} {024002} (\bibinfo {month}
  {Jul}\ \bibinfo {year} {2009}),\
  \Eprint{http://arxiv.org/abs/0812.0069}{arXiv:0812.0069},\
  \url{http://link.aps.org/doi/10.1103/PhysRevD.80.024002}%
  \bibAnnoteFile{NoStop}{F09}%
\bibitem{F11}%
  \BibitemOpen
  \bibfield{author}{%
  \bibinfo {author} {\bibfnamefont{M.}~\bibnamefont{Favata}},\ }%
  \bibfield{journal}{%
  \Doi{10.1103/PhysRevD.84.124013}{\bibinfo {journal} {Phys. Rev. D}}\ }%
  \textbf{\bibinfo {volume} {84}},\ \bibinfo {pages} {124013} (\bibinfo {month}
  {Dec}\ \bibinfo {year} {2011}),\
  \Eprint{http://arxiv.org/abs/1108.3121}{arXiv:1108.3121},\
  \url{http://link.aps.org/doi/10.1103/PhysRevD.84.124013}%
  \bibAnnoteFile{NoStop}{F11}%
\bibitem{BDEI05dr}%
  \BibitemOpen
  \bibfield{author}{%
  \bibinfo {author} {\bibfnamefont{L.}~\bibnamefont{Blanchet}}, \bibinfo
  {author} {\bibfnamefont{T.}~\bibnamefont{Damour}}, \bibinfo {author}
  {\bibfnamefont{G.}~\bibnamefont{Esposito-Far{\`e}se}},\ and\ \bibinfo
  {author} {\bibfnamefont{B.~R.}\ \bibnamefont{Iyer}},\ }%
  \bibfield{journal}{%
  \bibinfo {journal} {Phys. Rev. D}\ }%
  \textbf{\bibinfo {volume} {71}},\ \bibinfo {pages} {124004} (\bibinfo {year}
  {2005}),\ \Eprint{http://arxiv.org/abs/gr-qc/0503044}{gr-qc/0503044}%
  \bibAnnoteFile{NoStop}{BDEI05dr}%
\end{thebibliography}%

\end{document}